\documentclass[doublecolumn]{IEEEtran}
\usepackage{times}
\IEEEoverridecommandlockouts

\usepackage{cite}

\usepackage{multicol}
\usepackage[bookmarks=true]{hyperref}
\usepackage{xcolor}
\usepackage[final]{graphicx}
\usepackage{bbm}
\usepackage{tikz,pgf}
\usetikzlibrary{shapes}
\usetikzlibrary{patterns}
\usepackage{pgfplots}

\usepackage{standalone}

\usepackage{amsmath,amssymb,amsthm}
\usepackage{subcaption}
\usepackage{comment}

\usepackage{etoolbox}
\usepackage{algorithm}
\usepackage{algpseudocode}
\usepackage{enumitem}
\usepackage{cuted}
\allowdisplaybreaks

\newtheorem{theorem}{Theorem}
\newtheorem*{remark}{Remark}

\newtheorem{proposition}{Proposition}
\newtheorem{definition}{Definition}

\title{Cloud-Cluster Architecture for Detection in\\ Intermittently Connected Sensor Networks}

\author{
	Michal Yemini,
	Stephanie Gil,
	and
	Andrea J. Goldsmith
 \thanks{
	M.\ Yemini and A.\ J.\ Goldsmith are with the Electrical and Computer Engineering Department, Princeton University, Princeton, NJ, 08544 USA, e-mails: \texttt {myemini@princeton.edu;  goldsmith@princeton.edu}.	S.\ Gil is with the School of Engineering and Applied Sciences, Harvard University, Cambridge, MA, USA, \texttt{Email: sgil@seas.harvard.edu}.}
	\thanks{A summary of the results presented in this paper was presented in the  IEEE Global Communications Conference 2020 \cite{9348131}.}
	\thanks{This work was supported partially by the AFOSR award \#002484665 and partially by the ONR YIP grant \#N00014-21-1-2714 and the National Science Foundation CAREER Award CNS-2114733.}
 \thanks{\\
  \copyright 2022 IEEE. Personal use of this material is permitted. Permission
from IEEE must be obtained for all other uses, in any current or future
media, including reprinting/republishing this material for advertising or
promotional purposes, creating new collective works, for resale or
redistribution to servers or lists, or reuse of any copyrighted
component of this work in other works.
}
}

\begin{document}
\maketitle

\begin{abstract}
 We consider a centralized detection problem where sensors experience noisy measurements and intermittent connectivity to  a centralized fusion center. 
 The sensors
 collaborate locally within predefined sensor clusters and fuse their noisy sensor data to reach a common local estimate of the detected event in each cluster.  The connectivity of each sensor cluster is intermittent and depends on the available communication opportunities of the sensors to the fusion center. Upon receiving the  estimates from all the connected sensor clusters the fusion center fuses the received estimates to make a final determination regarding the occurrence of the event across the deployment area. We refer to this hybrid communication scheme as a \emph{cloud-cluster} architecture. We propose a method for optimizing the decision rule for each cluster and analyzing the expected detection performance resulting from our hybrid scheme.  Our method is tractable and addresses the high computational complexity caused by heterogeneous sensors' and clusters' detection quality, heterogeneity in their communication opportunities, and non-convexity of the loss function.
 Our analysis shows that clustering the sensors provides resilience to noise in the case of low sensor communication probability with the cloud. For larger clusters, a steep improvement in detection performance is possible even for a low communication probability by using our cloud-cluster architecture.     
\end{abstract}


\section{Introduction}

The next generation of wireless infrastructure enables cloud connectivity, and with it, powerful centralized decision making based on sensor data. However,  cloud connectivity of sensors cannot be guaranteed at all times, particularly for sensors operating over mmWave frequency bands (see \cite{5876482,5783993,6732923,6834753,7511572,8047278,8416684}) or in complex and potentially remote environments (see \cite{yasamin,pappas,zavlanos,gilIJRR}). Thus, a new paradigm that takes intermittent connectivity of sensors into account is needed. Currently, the analysis for sensor networks often assumes one of two architectures: i) a centralized architecture that is fully connected, or ii) a distributed architecture, such as peer-to-cloud, where connectivity is intermittent. 
In reality, a fully centralized case where sensors convey information directly to the cloud, a.k.a. fusion center (FC), is faulty since connectivity to it is intermittent. Alternatively, a distributed architecture where a sensor conveys its information directly to all of its neighbors to reach a common estimate distributively is not always feasible \cite{5978198} as  this also suffers from a long convergence time in large networks. Therefore, adopting either of these extremes can be problematic when the assumption of a continuously connected system is not practical, and alternatively, requiring fully distributed communication leads to an overly conservative system. 

The best way to fuse noisy data between sensors locally, and communicate this information to the cloud on an intermittent and sporadic  basis, optimizes the trade-off between accuracy and reliability of transmission. Failure to correctly consolidate noisy information will sacrifice accuracy. Nonetheless, requiring raw sensory data to be submitted over the cloud can lead to poor reliability due to sparse connectivity or high scheduling overhead due to a high connectivity requirement.
Thus, the question of how the communication infrastructure affects resilience to noise and the decision making abilities of sensors presents a \textit{knowledge gap} in our understanding of the vulnerability of multi-robot decision making systems in real world environments where communication links are unreliable. This work aims at closing this gap by developing an analytical framework to evaluate  the tradeoff between reliability of transmission and accuracy of estimation, as well as  provide system designs that are robust to acute link failures.

As network architectures evolve, multi-sensor systems operating in environments with limited connectivity may utilize a combination of centralized and distributed network architectures through a \emph{hybrid} local (i.e., clustered) network \emph{and} a (sporadically available) cloud network. We call this a \emph{cloud-cluster} communication architecture.  Such hybrid communication architectures give rise to important questions such as 1) \emph{how should the data be fused} at a local level in order to achieve the best global decision making ability at the cloud? and 2) what is the optimal size for the sensor clusters that would provide some \emph{resilience to sensor noise and sporadic connectivity of sensors to the cloud?}  Answering these questions would allow us the necessary insight to best optimize a cloud-cluster communication architecture for multi-sensor decision making.

This paper investigates the best architecture to achieve reliable prediction in the case of multiple sensors detecting an event of interest in the environment. In particular, we study a hybrid architecture where clusters of sensors pre-process their noisy observations, sending a compressed lower-dimensional aggregate observation to the cloud according to the probabilistic availability of the link. 
We develop a parameterized understanding of the trade-offs involved between architectures; either using larger clusters of sensors approaching a cluster-based (distributed) communication scheme, or using smaller clusters of sensors approaching a cloud based (centralized) communication scheme.  
We show that the cloud-cluster  architecture can drastically  improve resilience to noise when communication to the cloud is sporadic such as in real-world environments. We quantify the sensing noise of an individual sensor by its missed detection and false alarm probabilities, and  its intermittent connectivity to the cloud by a Bernoulli  random variable. 
Finally, we measure the prediction performance of the network architectures we consider in this work by the expected loss function, formally defined in \eqref{eq:error_probability_general2}. The expected loss function is  a linear combination of the false alarm and missed detection probabilities at the FC which captures the expected penalty caused by each of these detection errors.

\subsection{Paper Contributions}

In what follows, we highlight the  main contributions of our work: 
\begin{itemize}
    \item \textbf{Analyzing intermittent connectivity to the FC:} We present and formulate a model for sensor networks with intermittent connectivity to the FC. We propose to utilize a hybrid cloud-cluster communication architecture to overcome the harmful effect of the sensors' intermittent connectivity to the FC. To the best of our knowledge the case of intermittent connectivity to the FC, and the use and accompanying analysis of sensor clustering as a means to improve connectivity, has not previously been studied.
    \item \textbf{Exact and approximate solution for the homogeneous case:} We study and optimize a homogeneous system model where all sensors have the same sensing precision and probability of connectivity to the FC. For this case, the expected loss can be computed and minimized exactly. Additionally, we approximate the expected loss at the FC for this model when the number of sensors is large and discuss the resulting insights.
    \item \textbf{Approximate solution for the heterogeneous case:} In practical scenarios, sensing quality and connectivity to the FC for different sensors can differ, as well as the number of sensors in different clusters.  For these cases the exact error probability computation is intractable.  We propose an approximate solution for computing the false alarm and missed detection probabilities that factors in the randomly failed connections to the FC.
    Additionally, we use an iterative Gauss-Seidel method and a line search to optimize the cluster-level decision with the aim of minimizing the expected loss at the FC. 
    \item \textbf{Numerical results to support our analysis:} We present numerical results that support the theory developed in this paper.  Interestingly, these results show that clustering sensors creates a fundamental trade-off. On the one hand, clustering the sensors creates a lossy compression at the cluster level, and thus can increase the expected loss at the FC. On the other hand, the sensor's clustering can \textit{decrease} the false alarm and missed detection probabilities and the resulting expected loss at the FC since it increases the number of sensors that take part in the FC's decision when connectivity to the FC is poor.
\end{itemize}

\subsection{Related Work}

There has been much work in the area of determining analytical rules for event detection in clustered sensor networks. In particular, the works \cite{Cluster:4102537,Tsitsiklis93decentralizeddetection,Cluster:4407646,Tsitsiklis1988,Cluster:4608995,Cluster:4957097,Cluster:5751237,Cluster:EURASIP,Cluster:8649751,9095393} consider clustered sensor networks as a network organization scheme to reduce the communication overhead to the FC. Sensor networks are often characterized by extreme power and communication constraints and thus the objective in decentralized detection for these systems is to perform well, in their ability to detect an event, while transmitting the smallest number of bits possible. While these works make a significant contribution to our understanding of the clustered sensor networks, they do not consider the sporadic nature of the intermittent connectivity of multi-sensor systems. 
This aspect of the problem is very important, for example, in mmWave communication systems \cite{5876482,5783993,6732923,8416684}
that are vulnerable to temporary blockages, also known as outages. When a channel is blocked, no information can be passed through it, as its capacity is zero. These blockages occur with positive and non-negligible probability when the distance between a transmitter and receiver is greater than 150m, as is modeled in \cite{6834753,7511572,8047278}. Furthermore, they become more frequent as the distance between the transmitter and receiver grows. Connectivity is also a common problem  in mobile robotic systems (see \cite{yasamin,pappas,zavlanos,gilIJRR}), where robot location affects both the robot connectivity to the FC, and its event-detection probability.
To the best of our knowledge, minimizing the expected loss function of cloud-cluster sensor networks where sensors are intermittently connected to the cloud was not previously investigated. 
In this work we show that, using recently improved concentration inequalities, we can approximate the expected loss function caused by detection errors.   We note that like prior works \cite{Cluster:4102537,Tsitsiklis93decentralizeddetection,Cluster:4407646,Cluster:4608995,Cluster:4957097,Cluster:5751237,Cluster:EURASIP,Cluster:8649751}, we do not address the problem of optimizing sensor placement, or how to cluster existing sensors, but rather analyze the performance of existing system architectures.

Another related body of works analyzes the effect of the communication channel on the detection performance \cite{1386538,1597566,5992836,6225392,6678306}. 
These works study the effect of the quality of the communication channel, available side information and transmission power constraints on the distortion of the signals that are sent to the FC by the sensors. Our work considers a starkly different setup where channels from sensors to the FC may be blocked, thereby causing intermittent connectivity. In this case,  no information can be received by the FC from sensors with blocked channels. Our system architecture aims at improving connectivity to the FC using sensor clustering with optimized decision rules.

\paragraph*{Paper Organization}
The rest of the paper is organized as follows: Section \ref{sec:system_model} presents the system model and problem formulation. Section \ref{sec:analysis} analyzes  the optimal cloud-cluster decision rules. Sections \ref{sec:approx_prob_homogenous} and \ref{sec:approx_prob_heterogeneous}   include approximations to the optimal decision rules when they are intractable. In particular Section \ref{sec:approx_prob_homogenous} presents 
system analysis and optimization for a homogeneous system setup, whereas 
Section \ref{sec:approx_prob_heterogeneous} includes tractable analysis and decision rules  for heterogeneous setups. Section \ref{sec:numerical_results} presents numerical results. Finally, Section \ref{sec:conclusion} concludes the paper.

\section{System Model and Problem Formulation}\label{sec:system_model}
 This section presents the  system model this work studies  and the technical details of the associated system optimization problem.

\subsection{System Model}
Denote $[1:N]\triangleq \{1,\ldots,N\}$. We consider a set of sensors indexed by $i$, $i\in[1:N]$, that are deployed to sense the environment and determine if the event of interest has occurred.  We assume that the sensors are noisy and their ability to detect the event is captured for sensor $i$, by the probabilities $P_{\text{MD},s_i}$ of missed detection  and $P_{\text{FA},s_i}$ of false alarm.  Suppose that there are two hypotheses
$\mathcal H_0$ and $\mathcal H_1$, the first occurs with probability $P_0= 1-P_1$ and the second with probability $P_1$. We denote the random variable that symbolizes the correct hypothesis by $\Xi$, where $\Xi\in\{0,1\}$.  We assume for each sensor $i$ that the measured bit $y_i$ may be swapped with the following probabilities
\begin{flalign*}
 P_{\text{FA},s_i}&\triangleq \Pr(y_i=1|\Xi=0),\nonumber\\
P_{\text{MD},s_i} &\triangleq \Pr(y_i=0|\Xi=1),
\end{flalign*} 
where $P_{\text{FA},s_i},P_{\text{MD},s_i}\in (0,0.5)$ without loss of generality. We allow for heterogeneity in each sensor's ability to detect the event of interest. In practice these can arise due to characteristics such as the quality of their sensors and their proximity to the measured event.  
The sensors have intermittent connectivity to a centralized cloud server, or FC. This intermittent connectivity is modeled by a binary random variable $t_i$ that is equal to $1$ if sensor $s_i$ can communicate with the FC and $0$ otherwise. We denote by $p_{\text{com},s_i}$ the probability that sensor $s_i$ can communicate with the cloud (or FC), that is, $p_{\text{com},s_i}=\Pr(t_i=1)$. Upon obtaining a communication link to the cloud server, a \textit{communicating} sensor will  transmit information to the FC. The FC gathers the information it receives from the communicating sensors, 
and aims at estimating the correct hypothesis by minimizing the following expected loss function:
\begin{flalign}\label{eq:error_probability_general2}
E(L) \triangleq \Pr(\Xi=0)P_{\text{FA}}L_{10}+\Pr(\Xi=1)P_{\text{MD}}L_{01},
\end{flalign}
where $L_{10}$ is the loss caused by false alarm, $L_{01}$ is the loss caused by missed detection. Additionally, $P_{\text{FA}}$ and $P_{\text{MD}}$ are the false alarm and missed detection probabilities resulting from the  FC detection decision, respectively. Next, we present the three system communication architectures we consider in this work.

\subsection{System Communication Architecture}
\begin{figure}[t!]
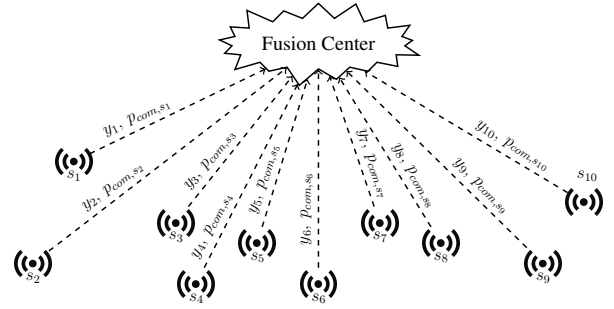

	\centering
	\includestandalone[mode=buildnew,scale=0.7]{system_setup_cloud}
	\caption{Sensor network with cloud architecture.}
	\label{fig:architecture_cloud}
\end{figure}

We consider three system communication architectures, namely, the cloud architecture, the cluster architecture and cloud-cluster hybrid architecture that generalizes the two aforementioned models. Next, we define each of these architectures.
 \begin{definition}[Cloud Architecture]
In a cloud architecture, see Fig.~\ref{fig:architecture_cloud}, all the sensors transmit their sensor data, $y_i$, to the cloud whenever a communication opportunity to the cloud exists.  Connectivity to the cloud is provided as a probability $p_{\text{com},s_i}$. The favorable case that $p_{\text{com},s_i}=1$ for all $i$ is equivalent to the classical centralized case since here all sensors have constant access to the cloud which in turn has access to all sensed measurements for event detection.
\end{definition}

In the case of constant sensor connectivity to the FC the cloud architecture minimizes the expected loss at the FC. However, in many realistic scenarios the FC  suffers from loss of connectivity to many sensors when connectivity is low. This  drastically increases the detection error probabilities and the resulting expected loss at the FC. 
 We propose an alternative approach aiming to improve network connectivity to the FC when sensor's connectivity to the FC is poor to minimize the expected loss at the FC. 

We study a different communication architecture where the sensors in the system are clustered into teams, and the sensors in each of these teams communicate with one another to arrive at a joint decision. This decision is then forwarded to the FC by a member of the cluster that can communicate with the FC. In this way, a cluster's decision can be forwarded to the FC if at least one sensor in the cluster can communicate with the FC.  Upon receiving the processed measurement from the clusters, the FC estimates the correct hypothesis by minimizing \eqref{eq:error_probability_general2} over all sensor clusters.
We call this hybrid design of the sensor
 communication architecture a cloud-cluster architecture. 

\begin{figure}[t!]
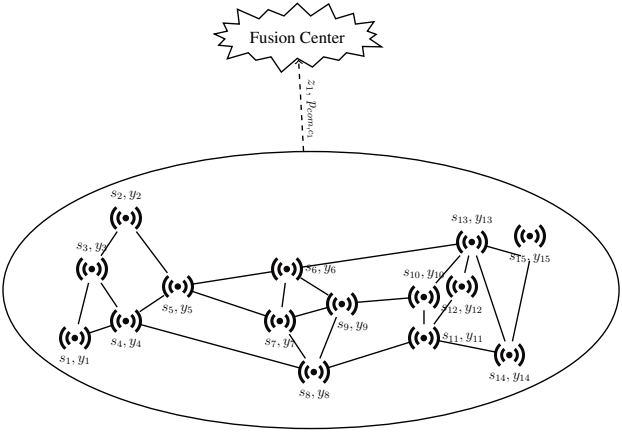

	\centering
	\includestandalone[mode=buildnew,scale=0.65]{system_setup_cluster}
	\caption{Sensor network with cluster architecture.}
	\label{fig:architecture_cluster}
\end{figure}

\begin{figure}[t!]
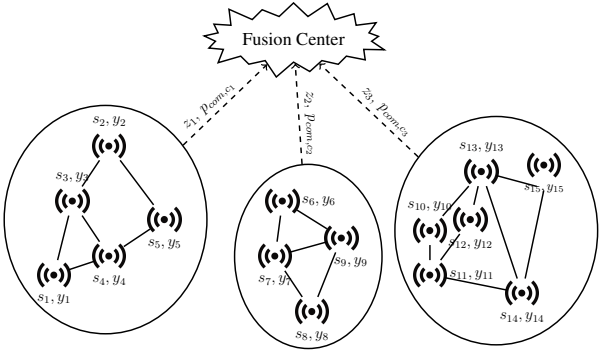

	\centering
	\includestandalone[mode=buildnew,scale=0.7]{system_setup_clusters_FC}
	\caption{Sensor network with cloud-cluster architecture.}
	\label{fig:architecture_cloud_cluster}
\end{figure}

\begin{definition}[Cluster Architecture]
In a cluster architecture, depicted in Fig.~\ref{fig:architecture_cluster}, all sensors have a fully connected local network and form a cluster where data is fused at a local level before being transmitted to the cloud.  Connectivity to the cloud exists if any sensor $s_i$ can communicate with the cloud. In this case, the fused sensor data is transmitted to the cloud by the sensor $s_i$. 
\end{definition}
\begin{definition}[Cloud-cluster Architecture]
The cloud-cluster architecture, depicted in Fig.~\ref{fig:architecture_cloud_cluster}, is a hybrid between a \emph{cloud} and a \emph{cluster} architecture where sensors are divided into several \emph{clusters}. It is assumed that sensors within a cluster are fully connected and can communicate locally.  The number of clusters in the system can range from 1 (cluster architecture)  to $N$ (cloud architecture) and is often determined by the problem settings, i.e. sensors operating in the same room of a building would constitute a cluster. Sensed data by sensors operating in a cluster is fused at a local level before being transmitted to the cloud. Connectivity to the cloud exists for each cluster if there is a sensor in the cluster that can communicate with the cloud.  In this case the fused sensor data for that cluster is transmitted to the cloud.
\end{definition}

\paragraph*{Intra-cluster Connectivity}
We note that this work focuses on sensor's intermittent connectivity to the FC and ignores intermittent communication links within a cluster. In practice, the communication requirements within a cluster are much less restrictive. In fact, there is no need for \textit{all} the sensors in a cluster to be directly connected to one another, or with a designated head sensor. Instead, it is sufficient to have a path of connected sensors between any two sensors in the cluster.
This occurs in a mmWave channel if every two consecutive sensors in the path are within a distance of 150m of each other (see \cite{6732923}).  Furthermore, even when communication between neighboring sensors in the path is intermittent, the work \cite{6732923} establishes that the blockage probability of the mmWave channel connecting them is considerably lower than their blockage probabilities to the faraway FC. Thus, for simplicity of exposition, this work focuses on the failing links between  sensors and the FC.

\paragraph*{Carrying Out Cluster Decisions}
Our cloud-cluster model is not limited to a specific decision making procedure within a cluster. 
For example, the cluster's decision rule can be performed by choosing  a designating cluster head to make a decision, this cluster head need not be fixed and can be chosen to maximize sensors' battery life \cite{5416081,s19194060}. The designated cluster head can forward the cluster's decision directly to the FC if it has a communication opportunity to it, or relay it to sensors in the cluster with a communication opportunity to the FC.
Alternatively, the  cluster's decision rule can be performed distributively. We note that as the number of sensors in a cluster grows, the latency of these procedures grows as well. Thus, in practical systems, the number of sensors in a cluster will be affected by the latency values that can be tolerated.   

We aim at analytically studying the performance of each model as a function of probability of connectivity to the cloud $p_{\text{com},s_i}$, and sensor noise which is captured by the probabilities $P_{\text{FA},s_i},P_{\text{MD},s_i}$.
Since the cloud architecture and the cluster architecture are special cases of the cloud-cluster architecture, our analysis is presented for the case of a cloud-cluster architecture.

\subsection{The Intermittently Connected Cloud-Cluster Problem Formulation}
We consider a hybrid cloud-cluster system comprising of $N_c$ clusters, denoted by $\mathcal{C}_1,\ldots,\mathcal{C}_{N_c}$.
\begin{definition}[Cluster connectivity]\label{def:cluster_connectivity}
A cluster $\mathcal{C}_j$  communicates with the FC if at least one of the sensors within the cluster
 can communicate with the FC. 
 \end{definition}
 Let $\tau_j$ be a binary random variable that is equal to one if \textit{cluster} $\mathcal{C}_j$ is communicating with the FC and zero otherwise and  denote   $\boldsymbol{\tau}=(\tau_1,\ldots, \tau_{N_c})$.
 
  Every sensor cluster $\mathcal{C}_j$ communicating with the cloud sends a pre-processed value $z_j$ that captures the  observations of all sensors in cluster $j$. If cluster $\mathcal{C}_j$ cannot communicate with the FC $z_j$ will take an arbitrary predefined deterministic value.
  We denote the vector of the pre-processed values by $\boldsymbol{z}=(z_1,\ldots, z_{N_c})$. The FC at the cloud determines its final decision of whether an event has occurred or not by using the optimal decision rule to minimize~\eqref{eq:error_probability_general2}. 
It follows from \cite[Chapter 3]{Kay:1993:FSS:151045} that this optimal decision rule  chooses  hypothesis $\mathcal{H}_1$ if:
\begin{flalign}\label{eq:dec_rule_basic}
\frac{\Pr(\boldsymbol z|\mathcal{H}_1,\boldsymbol{\tau})}{\Pr(\boldsymbol z|\mathcal{H}_0,\boldsymbol{\tau})} \geq \frac{L_{10}P_0}{L_{01}P_1}
\end{flalign}
and $\mathcal{H}_0$ otherwise.

We investigate the following questions: 
\begin{enumerate}
\item how the data $\boldsymbol{z}$ is pre-processed at the cluster layer to reduce the expected loss at the FC, 
\item how the estimates of missed detection and false alarm  probabilities are impacted by the system architecture, i.e., the number of clusters and the number of sensors per cluster,
\item how intermittent communication with the cloud impacts the performance at the FC which is captured by its  expected loss function.
\end{enumerate}

\section{System Analysis and Optimization}\label{sec:analysis}
In this section, we optimize the decision at  the cluster level and the FC. Additionally, we obtain the expected number of clusters that can communicate with the FC under the cloud-cluster architecture.

\subsection{Cloud-Cluster Communication}

Our cloud-cluster architecture  is aimed at
improving connectivity to FC when the probabilities $p_{\text{com},s_{i}}$ are small, and reducing scheduling and communication overheads when the probabilities $p_{\text{com},s_{i}}$ approach $1$.  
We assume that the sensors are clustered into $N_c$ groups.  As stated in Definition \ref{def:cluster_connectivity}, a cluster of sensors communicates with the FC if one of the sensors comprising the cluster  sees a communication opportunity to the FC. Each cluster estimates the hypothesis and sends its estimation to the FC provided there is a communication opportunity to the FC.

\subsection{Communication probability of clusters and the expected number of communicating clusters}

By Definition \ref{def:cluster_connectivity}, the probability that the cluster $\mathcal{C}_j$ can communicate with the FC, i.e., $\tau_j=1$, is:
\begin{flalign}\label{eq:cluster_comm_prob}
p_{\text{com},\mathcal{C}_j} \triangleq 1-\prod_{i:s_i\in\mathcal{C}_j}(1-p_{\text{com},s_{i}}).
\end{flalign}
Let $n_{\mathcal{C}_j}$ be the number of sensors in cluster $\mathcal{C}_j$.
We can see that as we increase the number of sensors to the clusters, $p_{\text{com},\mathcal{C}_j}$ increases. Therefore, $p_{\text{com},\mathcal{C}_j}$ is maximized in the cluster architecture where $n_{\mathcal{C}_j}=N$. On the other hand, $p_{\text{com},\mathcal{C}_j}$ is minimized in the cloud architecture where $n_{\mathcal{C}_j}=1$.  Additionally, as we increase the probability that a sensor can communicate with the FC, $p_{\text{com},\mathcal{C}_j}$ is increased.
Denote $(x_j)_{j=1}^N\triangleq(x_1,\ldots,x_N)$.
From \eqref{eq:cluster_comm_prob}  we can calculate the following expected number of communicating clusters:
\begin{flalign}\label{eq:exp_num_com_clusters}
\eta\left(N_c,(\mathcal{C}_j)_{j=1}^{N_c},(p_{\text{com},s_{i}})_{i=1}^{N}\right) \triangleq   N_c-\sum_{j=1}^{N_c}\prod_{i:s_i\in\mathcal{C}_j}(1-p_{\text{com},s_{i}}).
\end{flalign}
The optimization of the term \eqref{eq:exp_num_com_clusters} is beyond the scope of this paper since we assume a given clustering. Nonetheless, a closer look at the term  \eqref{eq:exp_num_com_clusters} provides the following key observations.
First, the expected number of communicating clusters is affected by three factors, namely, the number of clusters, the number of sensors in each cluster and the probability of connectivity to the FC. Second, the function $\eta$ is monotonically increasing with $p_{\text{com},s_{i}}$. However, the relationship between $N_c$, $|\mathcal{C}_j|$ and $\eta$ given a fixed number of sensors $N$ is more intriguing.  Considering, for example, the homogeneous case where $|\mathcal{C}_j|=N/N_c$ and $p_{\text{com},s_{i}}=p_{\text{com},s}$ we have that:
\[\eta=N_c\cdot\left(1-(1-p_{\text{com},s})^{N/N_c}\right)\]
Therefore, for small values of $p_{\text{com},s}$ decreasing the number of clusters $N_c$ increases $\eta$ instead of decreasing it; this behavior is observed until the probability $(1-p_{\text{com},s})^{N/N_c}$ becomes sufficiently small. When $p_{\text{com},s}$ is large, decreasing the number of clusters $N_c$ decreases $\eta$; in this scenario clustering reduces the scheduling overhead at the FC. 

\subsection{Decisions in Clusters}

While the objective in the FC is to minimize \eqref{eq:error_probability_general2} directly, the objective in the cluster level is to find the optimal trade-off between the probabilities of false alarm and missed detection. That is, the minimum probability of missed-detection that can be obtained for each value of the false alarm probability.
By the Neyman-Pearson Lemma \cite[Chapter 3]{Kay:1993:FSS:151045} the optimal trade-off can be found by using the following likelihood ratio test with a desired threshold $\gamma_j$: 
\begin{flalign}\label{eq:likelihoos_test_inside}
\frac{\Pr\left((y_i)_{i:s_i\in\mathcal{C}_j}|\mathcal{H}_1\right)}{\Pr\left((y_i)_{i:s_i\in\mathcal{C}_j}|\mathcal{H}_0\right)} \underset{\mathcal{H}_0}{\overset{\mathcal{H}_1}{\gtrless}} \gamma_j.
\end{flalign}
In  case of equality a random decision is made where hypothesis $\mathcal{H}_1$ is chosen with probability $p_j$ and hypothesis $\mathcal{H}_0$ is chosen with probability $1-p_j$, where $p_j$ is an additional parameter to be optimized.
Let 
\begin{align}
w_{1,s_{i}}&\triangleq\ln\left(\frac{1-P_{\text{MD},s_{i}}}{P_{\text{FA},s_{i}}}\right),\nonumber\\
w_{0,s_{i}}&\triangleq\ln\left(\frac{1-P_{\text{FA},s_{i}}}{P_{\text{MD},s_{i}}}\right),    
\end{align} 
and \[\tilde{y}_i\triangleq w_{1,s_{i}}y_i-w_{0,s_{i}}(1-y_i).\]
We can rewrite the likelihood ratio test \eqref{eq:likelihoos_test_inside} for decision in cluster $\mathcal{C}_j$  as follows:
\begin{flalign}
\sum_{i:s_i\in\mathcal{C}_{j}}\tilde{y}_i\underset{\mathcal{H}_0}{\overset{\mathcal{H}_1}{\gtrless}}\gamma_j.
\end{flalign} 
 In case of equality a random decision is made where hypothesis $\mathcal{H}_1$ is chosen with probability $p_j$ and hypothesis $\mathcal{H}_0$ is chosen with probability $1-p_j$.

Denote, 
\[P_{\text{FA},\mathcal{C}_j}\triangleq\Pr(z_j=1|\mathcal{H}_0), \quad \text{and}\quad P_{\text{MD},\mathcal{C}_j}\triangleq\Pr(z_j=0|\mathcal{H}_1).\] 
Then, the choice of threshold $\gamma_j$ and tiebreak probability $p_j$ results in the following detection error probabilities:
\begin{flalign}\label{P_errors_inside}
&P_{\text{FA},\mathcal{C}_j} = \Pr\left(\sum_{i:s_i\in\mathcal{C}_{j}}\tilde{y}_i> \gamma_j|\mathcal{H}_0\right) \nonumber\\
&\hspace{4cm}+ p_j\Pr\left(\sum_{i:s_i\in\mathcal{C}_{j}}\tilde{y}_i= \gamma_j|\mathcal{H}_0\right),\nonumber\\
&P_{\text{MD},\mathcal{C}_j} =\Pr\left(\sum_{i:s_i\in\mathcal{C}_{j}}\tilde{y}_i<\gamma_j|\mathcal{H}_1\right)\nonumber\\
&\hspace{2.5cm}+(1-p_j)\Pr\left(\sum_{i:s_i\in\mathcal{C}_{j}}\tilde{y}_i= \gamma_j|\mathcal{H}_1\right).
\end{flalign}
Generally, as we discuss in Section \ref{sec:threshol_opt_problem_general}, the calculation of the probabilities $P_{\text{FA},\mathcal{C}_j}$ and $P_{\text{MD},\mathcal{C}_j}$ is intractable except for special cases such as the homogeneous case analyzed in Section \ref{sec:approx_prob_homogenous}.
Therefore, our calculations for the general case, presented in  Section \ref{sec:approx_prob_heterogeneous}, rely on concentration inequalities to approximate $P_{\text{FA},\mathcal{C}_j}$ and $P_{\text{MD},\mathcal{C}_j}$. 

The threshold  $\gamma_j$ and the probability $p_j$ are parameters that we aim  at optimizing to reduce the expected loss at the FC for a given system architecture.
Denote 
\begin{align}
\ell_{\min,j} &\triangleq -\sum_{i:s_i\in\mathcal{C}_{j}}w_{0,s_{i}},\qquad
\ell_{\max,j} \triangleq \sum_{i:s_i\in\mathcal{C}_{j}}w_{1,s_{i}}.
\end{align}
The threshold $\gamma_j$ can be optimized by searching over the interval $\mathcal{L}_j=[\ell_{\min,j},\ell_{\max,j}]$ to minimize \eqref{eq:error_probability_general2}.  Additionally,  the probability $p_j$ can be optimized by searching over the interval $[0,1]$.
We note that the thresholds $\gamma_j$ and  probabilities $p_j$ that dictate the clusters' decisions do not depend on the set of clusters whose measurements are successfully received and fused at the FC, using the decision rule \eqref{eq:dec_rule_basic}. 
This choice obviates the need to  optimize  the thresholds $\gamma_j$ and the probabilities $p_j$  for all the possible $2^{N_c}$ combinations of communicating clusters. It also reduces the communication overhead that is caused by detecting the set of clusters that can communicate with the FC and sending this information back to the clusters for the correct choice of the  $\gamma_j$ and  $p_j$ every time the FC makes a detection decision.

\subsection{FC Final Decision}
Suppose that the cluster $\mathcal{C}_j$ is communicating with the FC and denote the data it sends to the FC by $z_i$. 
The optimal decision rule that minimizes  (\ref{eq:error_probability_general2}) is choosing hypothesis $\mathcal{H}_1$ whenever \eqref{eq:dec_rule_basic} holds
and hypothesis $\mathcal{H}_0$ otherwise. 
Let 
\begin{align}
w_{1,\mathcal{C}_j}&\triangleq\ln\left(\frac{1-P_{\text{MD},\mathcal{C}_j}}{P_{\text{FA},\mathcal{C}_j}}\right),\qquad
w_{0,\mathcal{C}_j}\triangleq\ln\left(\frac{1-P_{\text{FA},\mathcal{C}_j}}{P_{\text{MD},\mathcal{C}_j}}\right). 
\end{align}
The rule 
\eqref{eq:dec_rule_basic}
can be written as:
\begin{flalign*}
&\sum_{j=1}^{N_c}\tau_j\left[w_{1,\mathcal{C}_j}z_j-w_{0,\mathcal{C}_j}(1-z_j)\right]\geq\ln\left(\frac{L_{10}P_0}{L_{01}P_1}\right)\triangleq\gamma.
\end{flalign*}
Note that in the case of equality, the expected loss due to detection error is equal for both the false alarm and missed-detection errors. Thus, in the case of equality we may choose  hypothesis $\mathcal{H}_1$ arbitrarily since both hypotheses lead to the same loss. 

Thus, the sensing quality at the FC for a particular realization of the identity of communicating clusters can be written as
\begin{flalign*}
&P_{\text{FA}}(\boldsymbol{\tau}) =  \Pr\left(\sum_{j=1}^{N_c}\tau_j\left[w_{1,\mathcal{C}_j}z_j-w_{0,\mathcal{C}_j}(1-z_j)\right]\geq\gamma|\mathcal{H}_0,\boldsymbol{\tau}\right), \nonumber\\
&P_{\text{MD}}(\boldsymbol{\tau}) = \Pr\left(\sum_{j=1}^{N_c}\tau_j\left[w_{1,\mathcal{C}_j}z_j-w_{0,\mathcal{C}_j}(1-z_j)\right]<\gamma|\mathcal{H}_1,\boldsymbol{\tau}\right).
\end{flalign*}

The probability of that particular realization of the identity of communicating clusters is
\begin{flalign}
P(\boldsymbol \tau) = \prod_{j=1}^{N_c}p_{\text{com},\mathcal{C}_j}^{\tau_j}(1-p_{\text{com},\mathcal{C}_j})^{1-\tau_j}.
\end{flalign}
This results in the following sensing probabilities
\begin{flalign}\label{eq:error:prob_all_cluster2}
P_{\text{FA}} &=\Pr\left(\sum_{j=1}^{N_c}\tau_j\left[w_{1,\mathcal{C}_j}z_j-w_{0,\mathcal{C}_j}(1-z_j)\right]\geq\gamma|\mathcal{H}_0\right) \nonumber\\
&= \sum_{\boldsymbol \tau\in\{0,1\}^N}\hspace{-0.2cm}P(\boldsymbol \tau)P_{\text{FA}}(\boldsymbol \tau),\nonumber\\
P_{\text{MD}}&= \Pr\left(\sum_{j=1}^{N_c}\tau_j\left[w_{1,\mathcal{C}_j}z_j-w_{0,\mathcal{C}_j}(1-z_j)\right]<\gamma|\mathcal{H}_1\right) \nonumber\\ 
&= \sum_{\boldsymbol \tau\in\{0,1\}^N}\hspace{-0.2cm}P(\boldsymbol \tau)P_{\text{MD}}(\boldsymbol \tau). 
\end{flalign}

\subsection{The Threshold Optimization Problem}\label{sec:threshol_opt_problem_general}
Recall that $E(L)=Pr(\Xi=0)P_{\text{FA}}L_{10}+Pr(\Xi=1)P_{\text{MD}}L_{01}$ and that $P_{\text{FA}}$ and $P_{\text{MD}}$ are defined as \eqref{eq:error:prob_all_cluster2}.
Then, the global optimization problem resulting from the cloud-cluster architecture is:
\begin{align}\label{eq:overall_global_optimization_problem}
    &\min_{\{p_{j}\}_{j=1}^{N_c},\{\gamma_{j}\}_{j=1}^{N_c}} E(L).
\end{align}

The complexity of calculating the optimal values $p_j,\gamma_j$ is high for the following reasons: first,  the function $E(L)$ is not necessarily convex, thus the complexity can be exponential in the number of variables, i.e., exponential in $2N_c$. 
Additionally, currently no close form method is known to calculate \eqref{P_errors_inside} and \eqref{eq:error:prob_all_cluster2} efficiently since the coefficient are heterogeneous irrational numbers. We refer the reader to \cite{doi:10.1080/00949655.2018.1440294} for the case were the coefficients are rational numbers, additionally, the case of homogeneous coefficients is tractable as well. It follows that the overall complexity of optimizing $E(L)$ can be exponential in $2N_c+\max\{\max_j\{|\mathcal{C}_j|\},N_c\}$, where the last term in the addition follows from the calculation of \eqref{P_errors_inside} and \eqref{eq:error:prob_all_cluster2}.  

\section{Decision Optimization in Homogeneous Systems with Equal Thresholds}\label{sec:approx_prob_homogenous}
We consider a special case of our system model that is homogeneous, i.e., all the clusters comprises an equal number of homogeneous sensors where $P_{\text{FA},s_i}=P_{\text{FA},s}$, $P_{\text{MD},s_i}=P_{\text{MD},s}$, $P_{\text{com},s_i}=P_{\text{com},s},\: \forall\: i\in[1:N]$. 
In this case, $w_{1,s_i}=w_{1,s}$ and $w_{0,s_i}=w_{0,s}$ for all $i\in[1:N]$. 
For this setup,  we consider equal thresholds $\gamma_j$ and  probabilities $p_j$ of the clusters, i.e., $\gamma_j=\tilde{\gamma}_{\mathcal{C}}$ and $p_j=p_{\mathcal{C}}, \: \forall\: i\in[1:N]$. This leads to the tractability of Algo.~\ref{algo:homonegenous_setup_thoreshold}, at the expense of its optimality.

\subsection{Exact Optimization of the Expected Loss Function}
Recall that $P_{\text{FA},s_i},P_{\text{MD},s_i}\in (0,0.5)$, therefore, $w_{0,s}>0$ and $w_{1,s}>0$, and denote 
\begin{flalign*}
w_{1,s}&\triangleq\ln\left(\frac{1-P_{\text{MD},s}}{P_{\text{FA},s}}\right),\quad w_{0,s}\triangleq\ln\left(\frac{1-P_{\text{FA},s}}{P_{\text{MD},s}}\right),\nonumber\\
\gamma_{\mathcal{C}}&\triangleq\frac{\tilde{\gamma}_{\mathcal{C}}+|\mathcal{C}|\cdot w_{0,s}}{w_{1,s}+w_{0,s}}.
\end{flalign*} 
Under the assumptions of a homogeneous system and equal thresholds, we can rewrite \eqref{P_errors_inside} as
\begin{flalign}\label{P_errors_inside_homonesenous}
&P_{\text{FA},\mathcal{C}_j} 
=  \Pr\Bigg(\sum_{i:s_i\in\mathcal{C}_{j}}y_i> \gamma_{\mathcal{C}}|\mathcal{H}_0\Bigg)\nonumber\\
&\hspace{4cm}+p_{\mathcal{C}}\Pr\Bigg(\sum_{i:s_i\in\mathcal{C}_{j}}y_i= \gamma_{\mathcal{C}}|\mathcal{H}_0\Bigg),\nonumber\\
&P_{\text{MD},\mathcal{C}_j}=\Pr\Bigg(\sum_{i:s_i\in\mathcal{C}_{j}}y_i<\gamma_{\mathcal{C}}|\mathcal{H}_1\Bigg)\nonumber\\
&\hspace{1cm}+(1-p_{\mathcal{C}})\Pr\Bigg(\sum_{i:s_i\in\mathcal{C}_{j}}y_i=\gamma_{\mathcal{C}} |\mathcal{H}_1\Bigg).
\end{flalign}
 We can calculate the terms in \eqref{P_errors_inside_homonesenous}  efficiently for each $\gamma_{\mathcal{C}}$ since the  term $\sum_{i:s_i\in\mathcal{C}_j}y_i$ is distributed according to a binomial distribution for all $j\in[1:N_c]$.

The equal decision rules in the clusters create homogeneous clusters, i.e., $P_{\text{FA},\mathcal{C}_j}=P_{\text{FA},\mathcal{C}}$ and $P_{\text{MD},\mathcal{C}_j}=P_{\text{MD},\mathcal{C}}$ for all $j\in[1:N_c]$.
\textit{Hereafter, for simplicity of exposition, we assume in this section  that $P_{\text{FA},\mathcal{C}},P_{\text{MD},\mathcal{C}}<\frac{1}{2}$.} Our results can be easily extended to  the general case.
Denote
\begin{flalign*}
w_{1,\mathcal{C}}&\triangleq\ln\left(\frac{1-P_{\text{MD},\mathcal{C}}}{P_{\text{FA},\mathcal{C}}}\right),\quad w_{0,\mathcal{C}}\triangleq\ln\left(\frac{1-P_{\text{FA},\mathcal{C}}}{P_{\text{MD},\mathcal{C}}}\right),\nonumber\\
\gamma_{\text{h}}(k)&\triangleq\max\left\{0,\frac{\gamma+k\cdot w_{0,\mathcal{C}}}{w_{1,\mathcal{C}}+w_{0,\mathcal{C}}}\right\}.
\end{flalign*}

Let $\boldsymbol{1}$ denote the N-dimensional row vector with all entries equal to 1. Then, $P_{\text{FA}}(\boldsymbol{\tau}_1)=P_{\text{FA}}(\boldsymbol{\tau}_2)$ and $P_{\text{MD}}(\boldsymbol{\tau}_1)=P_{\text{MD}}(\boldsymbol{\tau}_2)$ for all $\boldsymbol{\tau}_1,\boldsymbol{\tau}_2\in\{0,1\}^{N_c}$ such that $\boldsymbol{\tau}_1\boldsymbol{1}^T=\boldsymbol{\tau}_2\boldsymbol{1}^T$ where $(\cdot)^T$ denotes the transpose operator. Additionally, denote
\begin{flalign*}
P_{\text{FA},k} &\triangleq  \Pr\Bigg(\sum_{j=1}^{k}z_i\geq \gamma_{\text{h}}(k)|\mathcal{H}_0,\boldsymbol{\tau}\boldsymbol{1}^T=k\Bigg),\nonumber\\ 
P_{\text{MD},k} &\triangleq \Pr\Bigg(\sum_{j=1}^{k}z_i<\gamma_{\text{h}}(k)|\mathcal{H}_1,\boldsymbol{\tau}\boldsymbol{1}^T=k\Bigg).
\end{flalign*}
Due to the homogeneity of the setup, the identity of the communicating clusters does not affect the probabilities $P_{\text{FA},k}$ and $P_{\text{MD},k}$.
Furthermore, by the homogeneity of the clusters, we have that 
\begin{flalign}
P_{\text{com},\mathcal{C}_j}=P_{\text{com},\mathcal{C}} = 1-\left(1-P_{\text{com},s}\right)^{N/N_c},
\end{flalign}
for all $j\in[1:N_c]$. 
Now, by
\eqref{eq:error:prob_all_cluster2}  for each pair $(p_{\mathcal{C}},\gamma_{\mathcal{C}})$ we have that
\begin{flalign}\label{eq:error_prob_hom_FA_MD_FC}
P_{\text{FA}} &= \sum_{k=0}^{N_c}\Pr\left(\boldsymbol{\tau}\boldsymbol{1}^T=k\right)P_{\text{FA},k}, \nonumber\\
P_{\text{MD}} &= \sum_{k=0}^{N_c}\Pr\left(\boldsymbol{\tau}\boldsymbol{1}^T=k\right)P_{\text{MD},k},
\end{flalign}
where $\boldsymbol{\tau}\boldsymbol{1}^T$ is a binomial random variable with $N_c$ experiments, each with probability of success $p_{\text{com},\mathcal{C}}$.
Therefore, the problem \eqref{eq:overall_global_optimization_problem}  can be upper bounded by 
\begin{align}\label{eq:overall_global_optimization_problem_homog}
    &\min_{p_{\mathcal{C}},\gamma_{\mathcal{C}}} E(L),
\end{align}
under the homogeneity assumptions included in this section. 

Recall that in a homogeneous setup all the clusters include an equal number of sensors. Therefore, the number of sensors in each cluster is $|\mathcal{C}|=N/N_c$.
Algo. \ref{algo:homonegenous_setup_thoreshold} depicts the resulting algorithm. It can be implemented with complexity of $O\left(r_p\frac{N}{N_c}\max\{\frac{N}{N_c}\ln^2(N/N_c),1\}\cdot\max\{N^2_c\ln^2(N_c),1\}\right)$, see \cite{PBinomia_dist_tailcomp} for efficient computation of the binomial tail distribution. 
It follows from \eqref{P_errors_inside_homonesenous} that the optimal value of $\gamma_{\mathcal{C}}$, under the homogeneity assumptions, is in the set $\{0,1,\ldots,N/N_c\}$. Additionally, we perform a line search in the interval $[0,1]$ to optimize the probability $p_{\mathcal{C}}$.

\begin{algorithm}[t!]
	\caption{Optimization for homogeneous setup and equal cluster thresholds  setup }\label{algo:homonegenous_setup_thoreshold}
		\begin{algorithmic}[1]		
		 \State Input: A set of $N_c$ homogeneous clusters $\mathcal{C}_1,\ldots,\mathcal{C}_{N_c}$, each comprises $|\mathcal{C}|=N/N_c$  homogeneous sensors;
		\State Input: $r_p\in\mathbb{N}_+$
		 \State Set $P_{\text{FA},s_i}=P_{\text{FA},s}$, $P_{\text{MD},s_i}=P_{\text{MD},s}$, $P_{\text{com},s_i}=P_{\text{com},s}$ for all $i\in[1:N]$;
		 		\State Set  $d_p=1/r_{p}$;
		\State Set $\Gamma_{\mathcal{C}} = \{0,1,\ldots,N/N_c\}$ and set $\Gamma_p=\{0,d_p,2d_p,\ldots,1\}$;
		 \State Set $P_{\text{FA},\mathcal{C}}$ and $P_{\text{MD},\mathcal{C}}$ as \eqref{P_errors_inside_homonesenous}.
		 \State Solve 
		$(p_{\mathcal{C}},\gamma_{\mathcal{C}})=\arg\min_{p_{\mathcal{C}\in \Gamma_p},\gamma_{\mathcal{C}}\in\Gamma_{\mathcal{C}}} E(L)$;
    \State Set $p_j=p_{\mathcal{C}}$ and $\gamma_j=\gamma_{\mathcal{C}}\cdot(w_{1,s}+w_{0,s})-|\mathcal{C}|\cdot w_{0,s}$ for all $j\in[1:N_c]$;
\end{algorithmic}
	\end{algorithm}

\subsection{Comparison of the Expected Loss for the Different Architectures}

The purpose of this section is to compare the minimal expected loss achievable by the different communication architectures, namely, the cloud-cluster,  cloud, and  cluster architectures. We express the  expected loss at the FC of each of these architectures as a function of the sensor's probability of communicating with the FC, which is a principle consideration of this paper.
We start with the first case of the cloud-cluster architecture.
Motivated by \cite{Tsitsiklis1988,Cluster:4608995,Cluster:4957097}, we  consider the case where $N_c,N\gg1$. Additionally, for the sake of simplicity of exposition we assume that $\gamma\geq0$ and $\gamma(k)/k\in(P_{\text{FA,s}},1-P_{\text{MD,s}}),\:\forall k\geq1$.  Let 
$D(p \parallel q)\triangleq p\ln(\frac{p}{q})+(1-p)\ln(\frac{1-p}{1-q})$,
denote the Kullback–Leibler (KL)  divergence. Recall that $P_{\text{FA},s},P_{\text{MD},s}<\frac{1}{2}$. Then, by  the Chernoff bound (see \cite{ARRATIA1989 }) and Stirling's formula (see \cite[p.115]{ash1990information}) we have that for every $\delta_{\mathcal{C}}\triangleq\gamma_{\mathcal{C}}\cdot \frac{N_c}{N}\in \left[P_{\text{FA},s},(1-P_{\text{MD},s})\right]$, 
\begin{flalign}\label{eq:approx_hom_cluster_errors} 
    \frac{1}{\sqrt{2N/N_c}}e^{-\frac{N}{N_c}D\left(\delta_{\mathcal{C}}\parallel P_{\text{FA},s}\right)}&\leq P_{\text{FA},\mathcal{C}}\leq e^{-\frac{N}{N_c}D\left(\delta_{\mathcal{C}}\parallel P_{\text{FA},s}\right)}, \nonumber\\
   \frac{1}{\sqrt{2N/N_c}}e^{-\frac{N}{N_c}D\left(\delta_{\mathcal{C}}\parallel 1-P_{\text{MD},s}\right)}&\leq  P_{\text{MD},\mathcal{C}}\leq e^{-\frac{N}{N_c}D\left(\delta_{\mathcal{C}}\parallel 1-P_{\text{MD},s}\right)}.
\end{flalign}
With a slight abuse of notations, we treat $\delta_{\mathcal{C}}$ as a real number when upper bounding the error probabilities since it can get  arbitrarily close to any real number as $N/N_c$ increases. A similar argument holds for the parameter $\beta$ which we define next.

\begin{figure*}
\begin{flalign}\label{eq:approx_hom_FC_errors}
   &E_{\text{cloud-cluster}}(L)\leq \inf_{\beta\in\left(\left(1-P_{\text{com},s}\right)^{\frac{N}{N_c}},1\right)}   \Bigg\{
   E_{\text{M}}\cdot\Pr\left(\boldsymbol{\tau}\boldsymbol{1}^T\leq (1-\beta) N_c\right)+\left(1-\Pr\left(\boldsymbol{\tau}\boldsymbol{1}^T\leq (1-\beta) N_c\right)\right)\cdot\nonumber\\
    &\hspace{4cm}\min_{\delta_{\mathcal{C}}\in \left[P_{\text{FA},s},(1-P_{\text{MD},s})\right]} \max_{k\geq(1-\beta)N_c}\Big\{(P_0L_{10}P_{\text{FA},k}+P_1L_{01}P_{\text{MD},k}\Big\}
    \Bigg\}\nonumber\\
    &\leq  \inf_{\beta\in\left(\left(1-P_{\text{com},s}\right)^{\frac{N}{N_c}},1\right)}   \Bigg\{E_{\text{M}}\cdot e^{-N_c D\left(\beta\parallel \left(1-P_{\text{com},s}\right)^{N/N_c}\right)}+\left(1-\frac{1}{\sqrt{2N_c}}e^{-N_c D\left(\beta\parallel \left(1-P_{\text{com},s}\right)^{N/N_c}\right)}\right)\cdot \nonumber\\
   &\hspace{0.5cm}\inf_{\delta_{\mathcal{C}}\in \left(P_{\text{FA},s},1-P_{\text{MD},s}\right):\:
  \frac{\gamma_{\text{h}}((1-\beta)N_c)}{(1-\beta)N_c},\frac{\gamma_{\text{h}}(N_c)}{N_c}\in \Gamma_{{\text{h}}}} \Big\{P_0L_{10}e^{-(1-\beta)N_c\cdot D\left(\frac{\gamma_{\text{h}}(N_c)}{N_c}\parallel P_{\text{FA},\mathcal{C}}\right)}+P_1L_{01}e^{-(1-\beta)N_c \cdot D\left(\frac{\gamma_{\text{h}}((1-\beta)N_c)}{(1-\beta)N_c}\parallel 1-P_{\text{MD},\mathcal{C}}\right)}\Big\}\Bigg\}.
\end{flalign}
\rule[-1ex]{2\columnwidth}{0.5pt}
\begin{flalign}\label{eq:approx_hom_FC_errors_cloud}
    &E_{\text{cloud}}(L)\geq \sup_{\beta\in\left(0,1-P_{\text{com},s}\right)} \Bigg\{ \nonumber\\  
    &\qquad\left(1-e^{-N \cdot D\left(\beta\parallel 1-P_{\text{com},s}\right)}\right)\cdot \frac{1}{\sqrt{2N(1-\beta)}}\Big[P_0L_{10}e^{-(1-\beta)N\cdot D\left(\gamma_{\text{h}}(1)\parallel P_{\text{FA},s}\right)}+P_1L_{01}e^{-(1-\beta)N \cdot D\left(\frac{\gamma_{\text{h}}((1-\beta)N)}{(1-\beta)N}\parallel 1-P_{\text{MD},s}\right)}\Big]
    \nonumber\\
    &\qquad+\frac{1}{2N}e^{-N\cdot D\left(\beta\parallel 1-P_{\text{com},s}\right)}\Big[P_0L_{10}e^{-N\cdot D\left(\frac{\gamma_{\text{h}}((1-\beta)N)}{(1-\beta)N}\parallel P_{\text{FA},s}\right)}+P_1L_{01}e^{-N \cdot D\left(\frac{\gamma_{\text{h}}(N)}{N}\parallel 1-P_{\text{MD},s}\right)}\Big]\Bigg\}.
\end{flalign}
\rule[-1ex]{2\columnwidth}{0.5pt}
\end{figure*}

 Let $\beta\geq \left(1-P_{\text{com},s}\right)^{N/N_c}$, then  the probability that more $\beta N_c$ clusters fail to communicate with the FC is bounded as follows
\begin{flalign}\label{eq:approx_disconnection_upper_home}
    \frac{1}{\sqrt{2N_c}}e^{-N_cD\left(\beta\parallel \left(1-P_{\text{com},s}\right)^{\frac{N}{N_c}}\right)}&\leq\Pr\left(\boldsymbol{\tau}\boldsymbol{1}^T\leq (1-\beta) N_c\right)\nonumber\\
    &\leq e^{-N_cD\left(\beta\parallel \left(1-P_{\text{com},s}\right)^{\frac{N}{N_c}}\right)}.
\end{flalign}
Denote $E_{\text{M}}\triangleq\min\{P_0L_{10},P_1L_{01}\}$, which is the minimal expected loss at the FC when no observations are available.
Utilizing the union bound results in the  upper bound \eqref{eq:approx_hom_FC_errors} for the  expected loss function at the FC for $\Gamma_{{\text{h}}}\triangleq \left(P_{\text{FA},\mathcal{C}},1-P_{\text{MD},\mathcal{C}}\right)$.

 Finally, we can plug-in the approximation \eqref{eq:approx_hom_cluster_errors} in \eqref{eq:approx_hom_FC_errors} to have a complete characterization on the trade-off between the parameters of the problem, namely the number of sensors $N$, the number of clusters $N_c$, the probability of a sensor's successful transmission to the FC $P_{\text{com},s}$ and its false alarm and missed detection probabilities, i.e., $P_{\text{FA},s}$ and $P_{\text{MD},s}$, respectively. We can observe from \eqref{eq:approx_hom_FC_errors} that for given $N$ and $N_c$ when $P_{\text{com},s} \downarrow 0$, i.e., links to the FC are very likely to disconnect, $\beta\uparrow 1$ and  $D\left(\beta\parallel \left(1-P_{\text{com},s}\right)^{\frac{N}{N_c}}\right)\downarrow0$. Additionally $(1-\beta)N_c\downarrow0$, therefore, we must decrease the number of clusters $N_c$ to decrease \eqref{eq:approx_hom_FC_errors}. Next, we consider how the number of clusters $N_c$ affects \eqref{eq:approx_hom_FC_errors} for a given $P_{\text{com},s}$. First, we observe that decreasing $N_c$ also decreases \eqref{eq:approx_hom_cluster_errors}. 
Nonetheless, the same claim is not necessarily true for  \eqref{eq:approx_hom_FC_errors}. Let us decrease $N_c$, then the probability of link failure from a cluster to the cloud, i.e., $\left(1-P_{\text{com},s}\right)^{N/N_c}$, decreases as well. However,  the exponents in \eqref{eq:approx_hom_FC_errors} also depend directly on the $N_c$ term that appears before the KL-divergence terms. Thus, in some cases, decreasing the number of clusters $N_c$ may result in increasing \eqref{eq:approx_hom_FC_errors} if the probability of cluster disconnection from the FC, which is captured by $\left(1-P_{\text{com},s}\right)^{N/N_c}$, and the precision of the decision at the cluster level, which is captured by \eqref{eq:approx_hom_cluster_errors} and is affected by $1-P_{\text{FA},s}$ and $1-P_{\text{MD},s}$, are not sufficiently increased with the decreasing of $N_c$. 
Section \ref{sec:numerical_results} which presents numerical results, include cases where decreasing $N_c$ results in increasing the expected loss. 

 To compare the performance of the cloud architecture, see Fig.~\ref{fig:architecture_cloud}, and the cloud-cluster architecture,  next we present a lower bound in the spirit of \eqref{eq:approx_hom_FC_errors} for the expected loss function of the cloud architecture, note that a similar upper bound can be directly derived from \eqref{eq:approx_hom_FC_errors} for $N_c=1$. Assuming that $N\gg1$ we observe that the events $\{\boldsymbol{\tau}\boldsymbol{1}^T\leq (1-\beta) N_c\}$ and $\{\boldsymbol{\tau}\boldsymbol{1}^T> (1-\beta) N_c\}$ are mutually exclusive with $\Pr\left(\boldsymbol{\tau}\boldsymbol{1}^T\leq (1-\beta) N_c\right)=1-\Pr\left(\boldsymbol{\tau}\boldsymbol{1}^T> (1-\beta) N_c\right)$. Thus, we can derive the lower bound \eqref{eq:approx_hom_FC_errors_cloud}. 

Following the discussion above, we observe that when the transmission probability $P_{\text{com},s}$ is small, the cloud-cluster architecture outperforms the cloud architecture. A special case that demonstrates the effectiveness of the cloud-cluster architecture is when $P_{\text{com},s}\propto \frac{1}{N}$. 
In this case $(1-\beta)\cdot N=O(1)$ when $\beta= 1-2P_{\text{com},s}$ as well as the expected loss $E_{\text{cloud}}(L)$.
This observation is of special interest in light of the works \cite{Tsitsiklis1988,Cluster:4608995,Cluster:4957097} which establish that when connectivity to the FC is perfect, clustering cannot reduce the expected loss function.

 We conclude this discussion by lower bounding in \eqref{eq:approx_hom_FC_errors_cluster} the expected loss function of the cluster architecture, depicted in Fig.~\ref{fig:architecture_cluster}. 
 A similar upper bound which excludes the factor $\frac{1}{\sqrt{2N}}$ in \eqref{eq:approx_hom_FC_errors_cluster} can be easily derived.
\begin{flalign}\label{eq:approx_hom_FC_errors_cluster}
    &E_{\text{cluster}}(L)\geq    E_{\text{M}} \left( 1-P_{\text{com},s}\right)^{N}+\frac{1-\left( 1-P_{\text{com},s}\right)^{N}}{\sqrt{2N}} \cdot\nonumber\\
&\Big[P_0L_{10}e^{-N\cdot D\left(\frac{\gamma_{\text{h}}((1-\beta)N)}{(1-\beta)N}\parallel P_{\text{FA},s}\right)}+P_1L_{01}e^{-N \cdot D\left(\frac{\gamma_{\text{h}}(N)}{N}\parallel 1-P_{\text{MD},s}\right)}\Big].
\end{flalign}
The probability that the FC does not receive transmissions from any of the clusters is $\left( 1-P_{\text{com},s}\right)^N$. Since this probability does not depend on the number of clusters,  the expected loss function $E(L)$ is minimized by the cluster architecture. Nonetheless, the cluster architecture cannot be used when the number of sensors is large due to the high latency it incurs.    
\begin{remark}
The upper bound \eqref{eq:approx_hom_FC_errors} is established by utilizing the union bound on the number of clusters that can communicate with the FC, and observing that due to the homogeneity assumption the identities of the communicating clusters do not affect \eqref{eq:approx_hom_FC_errors}. 
Therefore, this approach cannot be used in a heterogeneous setup, instead, in the following section, we develop an alternative large deviation approach that ties together a cluster's communication probability to the FC and its false alarm and missed detection probability.
\end{remark}

\section{Tractable Decision Optimization in Heterogeneous Systems}\label{sec:approx_prob_heterogeneous}
This section optimizes the decision thresholds for heterogeneous systems at the cluster level using the  Gauss-Seidel iterative method\footnote{The Gauss-Seidel iterative approach is considered in a relation to sensor network optimization in  \cite{Tsitsiklis93decentralizeddetection}. }  
which iteratively reduces the expected loss function at the FC. 
In the case that the terms  \eqref{P_errors_inside} and \eqref{eq:error:prob_all_cluster2} are intractable we approximate them via concentration inequalities.
Algo. \ref{algo:system_heterogeneous}  depicts the optimization scheme we develop in this section. Additionally, we propose several initial values for Algo. \ref{algo:system_heterogeneous} that we compare numerically in Section \ref{sec:numerical_results}.  
We note that for the sake of clarity of presentation we present  proofs and  analytical analysis   in Appendices \ref{append:primer_concentration}-\ref{append:proof_upper_FC_error_FA_MD}. Additionally, since Algo. \ref{algo:system_heterogeneous} does not minimize the expected loss exactly it may lead to suboptimal solutions.
Finally, hereafter we denote  $\{x_j\}_{j=1}^N\triangleq\{x_1,\ldots,x_N\}$. 
Finally, let $I_W$ be the number of iterations that are used in calculating the Lambert $W$ function \cite{lambert_w_funct_comp_complexity}. Then, Algo. \ref{algo:system_heterogeneous} is of complexity $O(T\cdot r_{\gamma}\cdot\max\{r_p2^{m_s},I_W+\frac{N}{N_c}\}\cdot\max\{2^{m_{\mathcal{C}}},I_W+N_c\})$.

\begin{algorithm*}
	\caption{Optimization for heterogeneous setup}\label{algo:system_heterogeneous}
\small
	\begin{algorithmic}[1]		
		\State Input: A set of clusters of sensors $\mathcal{C}_1,\ldots,\mathcal{C}_{N_c}$;
		\State Inputs: $\{\gamma_{j}^{(0)}\}_{j=1}^{N_c},\{p_{j}^{(0)}\}_{j=1}^{N_c}$;
		\State Inputs:
		$\{\ell_{\min,j}\}_{j=1}^{N_c}$, $\{\ell_{\max,j}\}_{j=1}^{N_c}$, and $r_{\gamma},r_p\in\mathbb{N}_+$;
		\State Inputs: $\overline{\delta}_{\gamma}>0$, $\overline{\delta}_{p}>0$, $T>0$,  $m_s>0$,$m_{\mathcal{C}}>0$;
		\State Set $\delta_{\gamma}^{(0)} = 2\overline{\delta}_{\gamma}$, $\delta_{p}^{(0)} = 2\overline{\delta}_{p}$, and $\Delta_{\gamma_j}=2\overline{\delta}_{\gamma}$ and $\Delta_{p_j}=2\overline{\delta}_{p}$ for all $j\in[1:N_c]$;
		\State Set $d_j=(\ell_{\max,j}-\ell_{\min,j})/r_{\gamma}$ for all $j\in[1:N_c]$ and set $d_p=1/r_{p}$;
		\State Set $\Gamma_j = \{\ell_{\min,j},\ell_{\min,j}+d_j,\ell_{\min,j}+2d_j,\ldots,\ell_{\max,j}\}$ and set $\Gamma_p=\{0,d_p,2d_p,\ldots,1\}$;
		\State Set  $t=0$, $j=0$, 
		\While {$t<T$}
		\While{$\delta_{\gamma}^{(t)}>\overline{\delta}_{\gamma}$ or $\delta_{p}^{(t)}>\overline{\delta}_{p}$}
		\State Set $t=t+1$;
		\State Set $j=\max\{\text{mod}(j+1,N_c),1\}$;
		\State Set $\gamma_k=\gamma_k^{(t-1)}$ and $p_k=p_k^{(t-1)}$ for all $k\in[1:N_c]$ such that $k\neq j$;
		 \If{$n_{\mathcal{C}_j}>m_s$ and $N_c>m_{\mathcal{C}}$}
		    \State Substitute $P_{\text{FA},\mathcal{C}_j}$ by its estimate  $U\left(n_{\mathcal{C}_j},\alpha_{\text{FA},j},M_{\text{FA},j},\sigma_{\text{FA},j}^2\right)$  in the calculation of $E(L)$.
		    \State Substitute $P_{\text{MD},\mathcal{C}_j}$ by its estimate   $U\left(n_{\mathcal{C}_j},\alpha_{\text{MD},j},M_{\text{MD},j},\sigma_{\text{MD},j}^2\right)$  in the calculation of $E(L)$.
		    \State Substitute  $P_{\text{FA}}$ by its estimate $U\left(N_c,\alpha_{\text{FA}},M_{\text{FA}},\sigma_{\text{FA}}^2\right)$  in the calculation of $E(L)$.
		    \State Substitute $P_{\text{MD},\mathcal{C}_j}$ by its estimate $U\left(N_c,\alpha_{\text{MD}},M_{\text{MD}},\sigma_{\text{MD}}^2\right)$ in the calculation of $E(L)$.
		    \State Set $p_j^{(t)}=1$ and  $\gamma_j^{(t)}=\min_{\gamma_j\in\Gamma_j} \overline{E}(L)$,
            where $\overline{E}(L)$ is calculated by using the estimation for the terms $P_{\text{FA},\mathcal{C}_j},P_{\text{MD},\mathcal{C}_j},P_{\text{FA}}$ and $P_{\text{MD}}$ in the calculation of $E(L)$;

		    \ElsIf{$n_{\mathcal{C}_j}\leq m_s$ and $N_c>m_{\mathcal{C}}$}
		    \State Substitute  $P_{\text{FA}}$ by its estimate $U\left(N_c,\alpha_{\text{FA}},M_{\text{FA}},\sigma_{\text{FA}}^2\right)$  in the calculation of $E(L)$.
		    \State Substitute $P_{\text{MD},\mathcal{C}_j}$ by its estimate $U\left(N_c,\alpha_{\text{MD}},M_{\text{MD}},\sigma_{\text{MD}}^2\right)$ in the calculation of $E(L)$.
		    
            \State Set $(\gamma_j^{(t)},p_j^{(t)})=\min_{\gamma_j\in\Gamma_j,p_j\in\Gamma_p} \overline{E}(L)$,
            where $\overline{E}(L)$ is calculated by using the estimation for the terms 
            $P_{\text{FA}}$ and $P_{\text{MD}}$ in the calculation of $E(L)$;

		    \ElsIf{$n_{\mathcal{C}_j}> m_s$ and $N_c\leq m_{\mathcal{C}}$}
		    \State Substitute $P_{\text{FA},\mathcal{C}_j}$ by its estimate  $U\left(n_{\mathcal{C}_j},\alpha_{\text{FA},j},M_{\text{FA},j},\sigma_{\text{FA},j}^2\right)$  in the calculation of $E(L)$.
		    \State Substitute $P_{\text{MD},\mathcal{C}_j}$ by its estimate   $U\left(n_{\mathcal{C}_j},\alpha_{\text{MD},j},M_{\text{MD},j},\sigma_{\text{MD},j}^2\right)$  in the calculation of $E(L)$.
		    \State Set $p_j^{(t)}=1$ and $\gamma_j^{(t)}=\min_{\gamma_j\in\Gamma_j} \overline{E}(L)$,
            where $\overline{E}(L)$ is calculated by using the estimation for the terms $P_{\text{FA},\mathcal{C}_j}$ and $P_{\text{MD},\mathcal{C}_j}$ in the calculation of $E(L)$;
		    \Else \State
		    Set $(\gamma_j^{(t)},p_j^{(t)})=\min_{\gamma_j\in\Gamma_j,p_j\in\Gamma_p} E(L)$;
		 \EndIf
		\State Set $\Delta_{\gamma_j} = |\gamma_j^{(t)}-\gamma_j^{(t-1)}|$ and set $\delta_{\gamma}^{(t)}=\max\{\Delta_{\gamma_k}\}_{k=1}^{N_c}$;
		\State Set $\Delta_{p_j} = |p_j^{(t)}-p_j^{(t-1)}|$ and set $\delta_{p}^{(t)}=\max\{\Delta_{p_k}\}_{k=1}^{N_c}$;
		\EndWhile 
		\EndWhile
	\end{algorithmic}
\end{algorithm*}

\subsection{From grid search to line search}
We overcome the non-convexity of the objective function of \eqref{eq:overall_global_optimization_problem} with respect to $\gamma_j$ and $p_j$ by optimizing these variables  using a combination of the Gauss-Seidel iterative method with a line search at each iteration. 
Starting from chosen initial values for $\gamma_j$ and $p_j$, this method optimizes the thresholds iteratively until convergence, one cluster at a time, while fixing the decision thresholds of all the other clusters. At each iteration a line search is performed over a predefined bounded interval to minimize the overall expected loss. We propose four different initial values for $\gamma_j$ and $p_j$ in Section \ref{sec:starting_points_het}.

\subsection{Approximating \eqref{P_errors_inside} and \eqref{eq:error:prob_all_cluster2} via concentration inequalities}\label{sec:concentration_clusters_FC}
Now, we explore optimizing the thresholds $\gamma_j$ via concentration inequalities, specifically, the improved Bennet's inequality that is stated in Theorem \ref{theorem:improved_bennett}, Appendix \ref{append:primer_concentration}.  We note that  it is possible to approximate the detection error probability  using the normal approximation. However, it  yields smaller approximate probabilities than the true ones, which we want to upper bound, when the false alarm and missed detection probabilities are small. Therefore, it is not suitable to use in the estimation of the loss function at the FC when the clusters are large.  Thus, for the clarity of presentation, we use the improved Bennet's inequality in our analysis, which upper bounds the desired probability in all scenarios. 

 Next, we present  the following notations. 
Additionally, let $W(\cdot)$ denote the Lambert $W$ function.  
Denote   
 \begin{flalign}\label{eq:U_def}
&U(n,\alpha,M,\sigma^2)\triangleq\nonumber\\
&\qquad\exp\left[-\frac{\Lambda \alpha}{M}+n\ln\left(1+\frac{\sigma^2}{M^2}\left(e^{\Lambda}-1-\Lambda\right)\right)\right],
\end{flalign}
where
	\begin{align}
	A &\triangleq \frac{M^2}{\sigma^2}+\frac{nM}{\alpha}-1,\nonumber\\
	B &\triangleq \frac{nM}{\alpha} -1,\nonumber\\
	\Lambda&\triangleq A-W(Be^A).
	\end{align}

We separate the concentration inequalities analysis into two scenarios, both of which are intractable on their own.

\subsubsection{Large number of sensors in cluster $j$ ($n_{\mathcal{C}_j}\gg1$)} In this case we approximate the false alarm and missed detection probabilities of the decision of cluster $j$ by applying the improved Bennet's inequality as follows.
\begin{proposition}\label{prop:upper_cluster_error_FA}
Let
\begin{align*}
\alpha_{\text{FA},j}&=\gamma_j-\sum_{i:s_i\in\mathcal{C}_{j}}(P_{\text{FA},s_{i}}w_{1,s_{i}}-(1-P_{\text{FA},s_{i}})w_{0,s_{i}}),\\
\sigma_{\text{FA},j}^2&=\frac{1}{n_{\mathcal{C}_j}}\sum_{i:s_i\in\mathcal{C}_{j}}P_{\text{FA},s_{i}}(1-P_{\text{FA},s_{i}})(w_{1,s_{i}}+w_{0,s_{i}})^2,
\end{align*}
and $M_{\text{FA},j} = \max_{i:s_i\in\mathcal{C}_{j}}\{m_{\text{FA},i}\}$ where
$m_{\text{FA},i} = (1-P_{\text{FA},s_{i}})(w_{1,s_{i}}+w_{0,s_{i}})$.
Then,
\begin{flalign}\label{eq:upper_cluster_error_FA}
P_{\text{FA},\mathcal{C}_j} \leq U\left(n_{\mathcal{C}_j},\alpha_{\text{FA},j},M_{\text{FA},j},\sigma_{\text{FA},j}^2\right),
\end{flalign}
for every $\gamma_j$ such that $0\leq \gamma_j-\sum_{i:s_i\in\mathcal{C}_{j}}(P_{\text{FA},s_{i}}w_{1,s_{i}}-(1-P_{\text{FA},s_{i}})w_{0,s_{i}})<n_{\mathcal{C}_j}\cdot M_{\text{FA},j}$.
\end{proposition}
\begin{proposition}\label{prop:upper_cluster_error_MD}
Denote
\begin{flalign*}
\alpha_{\text{MD},j}&=\sum_{i:s_i\in\mathcal{C}_{j}}((1-P_{\text{MD},s_{i}})w_{1,s_{i}}-P_{\text{MD},s_{i}}w_{0,s_{i}})-\gamma_j,\\
\sigma_{\text{MD},j}^2&=\frac{1}{n_{\mathcal{C}_j}}\sum_{i:s_i\in\mathcal{C}_{j}}P_{\text{MD},s_{i}}(1-P_{\text{MD},s_{i}})(w_{1,s_{i}}+w_{0,s_{i}})^2,
\end{flalign*}
and $M_{\text{MD},j} = \max_{i:s_i\in\mathcal{C}_{j}}\{m_{\text{MD},i}\}$ where
$m_{\text{MD},i} = (1-P_{\text{MD},s_{i}})(w_{1,s_{i}}+w_{0,s_{i}})$.
Then,
\begin{flalign}\label{eq:upper_cluster_error_MD}
P_{\text{MD},j} &\leq U\left(n_{\mathcal{C}_j},\alpha_{\text{MD},j},M_{\text{MD},j},\sigma_{\text{MD},j}^2\right),
\end{flalign}
for every $\gamma_j$ such that 
$0\leq \sum_{i:s_i\in\mathcal{C}_{j}}((1-P_{\text{MD},s_{i}})w_{1,s_{i}}-P_{\text{MD},s_{i}}w_{0,s_{i}})-\gamma_j<n_{\mathcal{C}_j}M_{\text{MD},j}$.

\end{proposition}	
We prove Proposition \ref{prop:upper_cluster_error_FA} and  Proposition \ref{prop:upper_cluster_error_MD} in Appendix \ref{append:proof_upper_cluster_error_FA_MD}.

\subsubsection{Large number of clusters ($N_{c}\gg 1$)} 
In this case we approximate the false alarm and missed detection probabilities of the decision of the FC by Propositions  \ref{prop:upper_FC_error_FA} and \ref{prop:upper_FC_error_MD}  that are achieved by applying  the improved Bennet's inequality. Interestingly, Propositions  \ref{prop:upper_FC_error_FA} and \ref{prop:upper_FC_error_MD} establish concentration inequalities that consider \textit{both} the detection error of a cluster and its probability of successful communication with the FC. 
\begin{proposition}\label{prop:upper_FC_error_FA}
Let 
\[E_{0,j} \triangleq p_{\text{com},\mathcal{C}_j}\left(P_{\text{FA},\mathcal{C}_{j}}w_{1,\mathcal{C}_{j}}-(1-P_{\text{FA},\mathcal{C}_{j}})w_{0,\mathcal{C}_{j}}\right),\] 
and denote
\begin{align*}
&\alpha_{\text{FA}}\triangleq\gamma-\sum_{j=1}^{N_c}E_{0,j},\nonumber\\
&\sigma_{\text{FA}}^2\triangleq\nonumber\\
&\frac{1}{N_c}\sum_{j=1}^{N_c}\Big[p_{\text{com},\mathcal{C}_j} \left(P_{\text{FA},\mathcal{C}_{j}}w_{1,\mathcal{C}_{j}}^2+(1-P_{\text{FA},\mathcal{C}_{j}})w_{0,\mathcal{C}_{j}}^2\right)  -E_{0,j}^2\Big],
 \end{align*}
 and $M_{\text{FA}} = \max_{j\in[1:N_c]}\{m_{\text{FA},j}\}$, where
$m_{\text{FA},j} =\max\{|w_{1,\mathcal{C}_j}-E_{0,j}|,|w_{0,\mathcal{C}_j}+E_{0,j}|\}$.
Then,
\begin{flalign}\label{eq:upper_FC_error_FA}
P_{\text{FA}} \leq U\left(N_c,\alpha_{\text{FA}},M_{\text{FA}},\sigma_{\text{FA}}^2\right),
\end{flalign}
for every $\gamma$ such that $0\leq \gamma-\sum_{j=1}^{N_c}E_{0,j}<N_c\cdot M_{\text{FA}}$.
\end{proposition}
\begin{proposition}\label{prop:upper_FC_error_MD}
Let 
\[E_{1,j} \triangleq p_{\text{com},\mathcal{C}_j}\left( (1-P_{\text{MD},\mathcal{C}_{j}})w_{1,\mathcal{C}_{j}}-P_{\text{MD},\mathcal{C}_{j}}w_{0,\mathcal{C}_{j}}\right),\] 
and denote
\begin{flalign*}
&\alpha_{\text{MD}}\triangleq\sum_{j=1}^{N_c}E_{1,j}-\gamma,\nonumber\\
&\sigma_{\text{MD}}^2\triangleq \nonumber\\
&\frac{1}{N_c}\sum_{j=1}^{N_c}\Big[p_{\text{com},\mathcal{C}_j} \left( (1-P_{\text{MD},\mathcal{C}_{j}})w_{1,\mathcal{C}_{j}}^2+P_{\text{MD},\mathcal{C}_{j}}w_{0,\mathcal{C}_{j}}^2\right)-E_{1,j}^2\Big]
\end{flalign*} 
and  $M_{\text{MD}} = \max_{j\in[1:N_c]}\{m_{\text{MD},j}\}$ where
$m_{\text{MD},j} =\max\{|w_{1,\mathcal{C}_j}-E_{1,j}|,|w_{0,\mathcal{C}_j}+E_{1,j}|\}$. Then,
\begin{flalign}\label{eq:upper_FC_error_MD}
P_{\text{MD}} &\leq U\left(N_c,\alpha_{\text{MD}},M_{\text{MD}},\sigma_{\text{MD}}^2\right),
\end{flalign}
for every $\gamma$ such that  
$0\leq \sum_{j=1}^{N_c}E_{1,j}-\gamma<N_c \cdot M_{\text{MD}}$.
\end{proposition}

We prove Propositions \ref{prop:upper_FC_error_FA} and  \ref{prop:upper_FC_error_MD} in Appendix \ref{append:proof_upper_FC_error_FA_MD}.

\begin{remark}
Propositions \ref{prop:upper_cluster_error_FA}-\ref{prop:upper_FC_error_MD} capture the heterogeneity of the model through the terms  $\alpha_{\text{FA},j}$,  $\alpha_{\text{MD},j}$, $\alpha_{\text{FA}}$ and $\alpha_{\text{MD}}$, and the variance terms $\sigma^2_{\text{FA},j},\sigma^2_{\text{MD},j},\sigma^2_{\text{FA}}$ and $\sigma^2_{\text{MD}}$. Furthermore, the network architecture is captured  by the number of clusters and the number of sensors at each cluster.
Additionally, the transmission probability of a cluster $j$ clearly affects the expectation terms $E_{0,j}$, $E_{1,j}$ and the variance terms $\sigma_{\text{FA}}^2$, and $\sigma_{\text{MD}}^2$ that are used to upper bound the false alarm and missed detection probabilities at the FC.
\end{remark}

Using  Propositions \ref{prop:upper_cluster_error_FA}-\ref{prop:upper_FC_error_MD} we can evaluate  and minimize the expected loss function to optimize the quality of detection even when the exact calculations are intractable. 

\subsection{Initial Inputs to Algorithm \ref{algo:system_heterogeneous}}\label{sec:starting_points_het}
 Since Algo. \ref{algo:system_heterogeneous} uses the Gauss-Seidel iterative algorithm it is required to provide it with the initial values $\{\gamma_{j}^{(0)}\}_{j=1}^{N_c},\{p_{j}^{(0)}\}_{j=1}^{N_c}$. 
We consider the following four initial values:
\begin{enumerate}
    \item For each cluster $\mathcal{C}_j$ the choice of $\gamma_{j}^{(0)}$ and $p_{j}^{(0)}$ is found using the equal threshold solution as in Algo. \ref{algo:homonegenous_setup_thoreshold} under the assumption that there are $N_c$ clusters that are identical to cluster $\mathcal{C}_j$, i.e. they include the same number of sensors as cluster $\mathcal{C}_j$ with the same probabilities of false alarm, missed-detection and communication to the cloud as the sensors in cluster $\mathcal{C}_j$. The probabilities $P_{\text{FA},\mathcal{C}}$ and $P_{\text{MD},\mathcal{C}}$ are calculated using the approximations we presented in Section \ref{sec:concentration_clusters_FC} if they are intractable.
    \item  Middle point of the intervals $\left[\ell_{\min,j},\ell_{\max,j}\right]$ and $[0,1]$, respectively. That is, 
    $\gamma_{j}^{(0)}=\frac{\ell_{\min,j}+\ell_{\max,j}}{2}$, and
 $p_{j}^{(0)}=0.5$.
    \item 
    $\gamma_{j}^{(0)}=\ell_{\min,j}$ and $p_{j}^{(0)}=1$, that is, $P_{\text{FA},\mathcal{C}_j}=1$, and $P_{\text{MD},\mathcal{C}_j}=0$.
    \item  $\gamma_{j}^{(0)}=\ell_{\max,j}$ and $p_{j}^{(0)}=0$, that is, $P_{\text{FA},\mathcal{C}_j}=0$, and $P_{\text{MD},\mathcal{C}_j}=1$.
\end{enumerate}

\section{Numerical Results}\label{sec:numerical_results}

This section presents numerical results in which we evaluate the performance of the proposed cloud-cluster architecture.
We consider a system with the following characteristics: 500 sensors, to evaluate both the actual and approximate performance, $p(\Xi=1) = 0.65$, $L_{01}=200$ and $L_{10}=100$. To evaluate the performance of the proposed approach we compare two systems: a homogeneous one in which $p_{\text{FA},s_i} = 0.2$, $p_{\text{MD},s_i} = 0.35$ for all the sensors in the network, and a heterogeneous system in which for each sensor $i$  we have that  $p_{\text{FA},s_i}\sim U([0.16,0.24])$ and $p_{\text{MD},s_i}\sim U([0.28,0.42])$, that is, both the false alarm and missed detection probabilities of each sensor has a random deviation of 20\% from their values in the homogeneous system. In the heterogeneous setup we average the expected loss of each realization of the false alarm and missed detection probabilities  over 250 realizations. Additionally, in each grid search that we perform for optimizing $\gamma_j$ we use $50$ points per sensor, i.e., a total of $r_{\gamma}=50\times n_{\mathcal{C}_j}$ points. Finally, the line search resolution for the  variable $p_j$ is $0.01$, that is, $r_p=100$.

\begin{figure}[t!]
    \centering
    \includegraphics[scale=0.6]{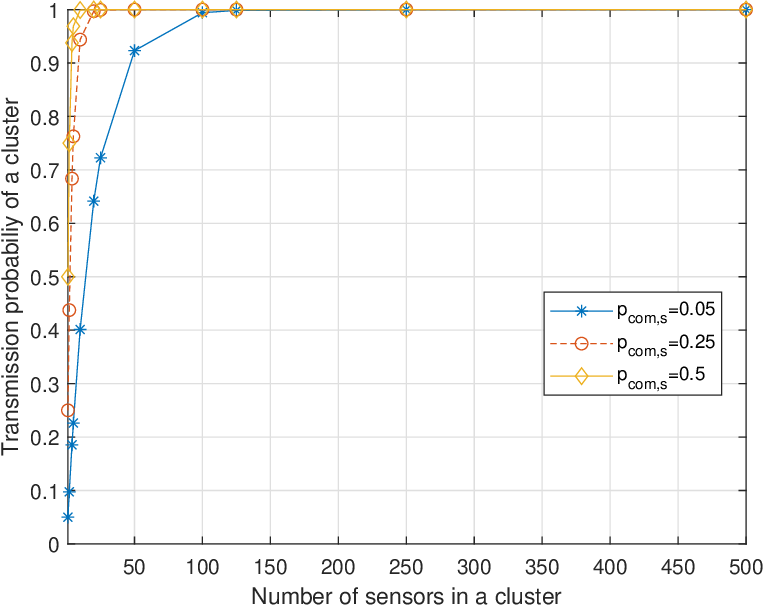}
    \caption{Communication probability to the cloud as a function of the number of sensors  it includes.}
    \label{fig:prob_comm_clusters}
\end{figure}

First, we evaluate in Fig. \ref{fig:prob_comm_clusters} the communication probability of a cluster to the cloud as a function of the  number of sensors it comprises for three values of individual sensor communication probability, $P_{\text{com},s_i}= 0.05, 0.25,0.5$. Fig. \ref{fig:prob_comm_clusters} validates that the communication probability of a cluster grows monotonically with the number of sensors it includes. Additionally, it shows that for higher  values of $P_{\text{com},s_i}$ the increase in communication probability occurs and saturates faster than for lower values of $P_{\text{com},s_i}$. 

\begin{figure}[t]
 \begin{subfigure}{0.5\textwidth}
 	\centering
	\includegraphics[scale=0.6]{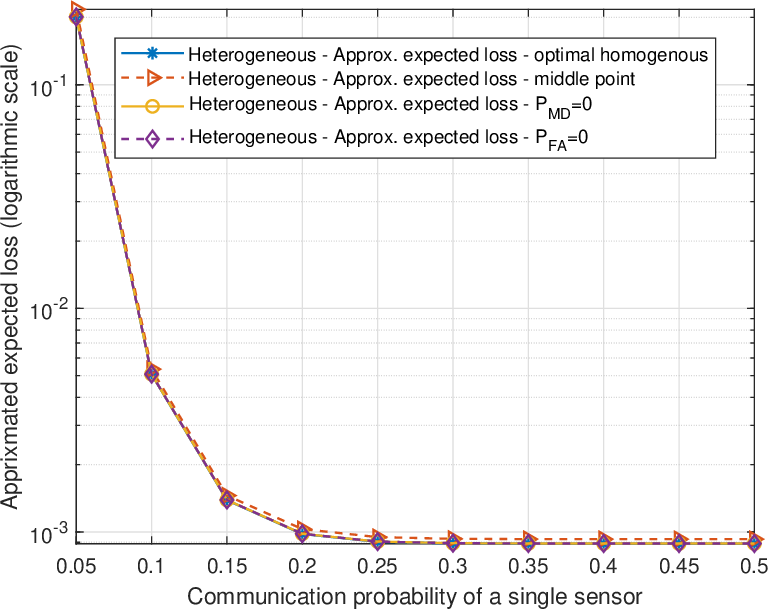}
	\caption{A system with $20$ clusters, each including $25$ sensors. }
	\label{Het_compare_init_fig_plot_N_c=10}
   \vspace{0.25cm}
 \end{subfigure}//
 \begin{subfigure}{0.5\textwidth}
 	\centering
	\includegraphics[scale=0.6]{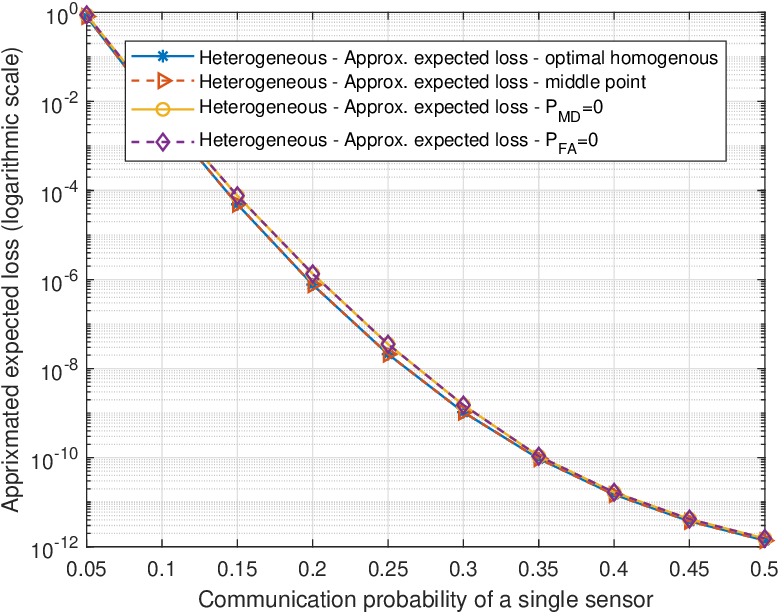}
	\caption{A system with $100$ clusters, each including $5$ sensors. }
	\label{Het_compare_init_fig_plot_N_c=50}
\end{subfigure}
\vspace{0.25cm}
\caption{The expected loss  as a function of the communication probability of each sensor, for each of the initial thresholds presented in Section \ref{sec:starting_points_het}. The approximated expected loss values resulting from the different initial thresholds are similar. Nevertheless, there is a small but persistent advantage for the  ``optimal homogeneous" initial threshold that minimizes the expected loss function assuming identical clusters.}
\label{fig_Het_compare_init_fig_plot_N_c}
 \end{figure}

\begin{figure}[t]
 \begin{subfigure}{0.5\textwidth}
 	\centering
	\includegraphics[scale=0.6]{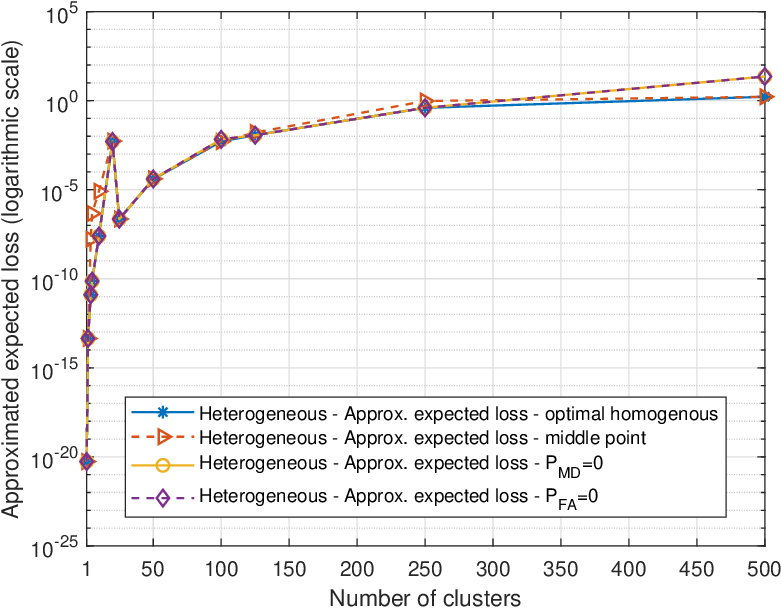}
	\caption{A system with sensor communication probability $p_{\text{com},s_i}=0.1$. }
	\label{Het_compare_init_fig_plot_p_c=0.1}
    \vspace{0.25cm}
 \end{subfigure}\\
\begin{subfigure}{0.5\textwidth}
 	\centering
	\includegraphics[scale=0.6]{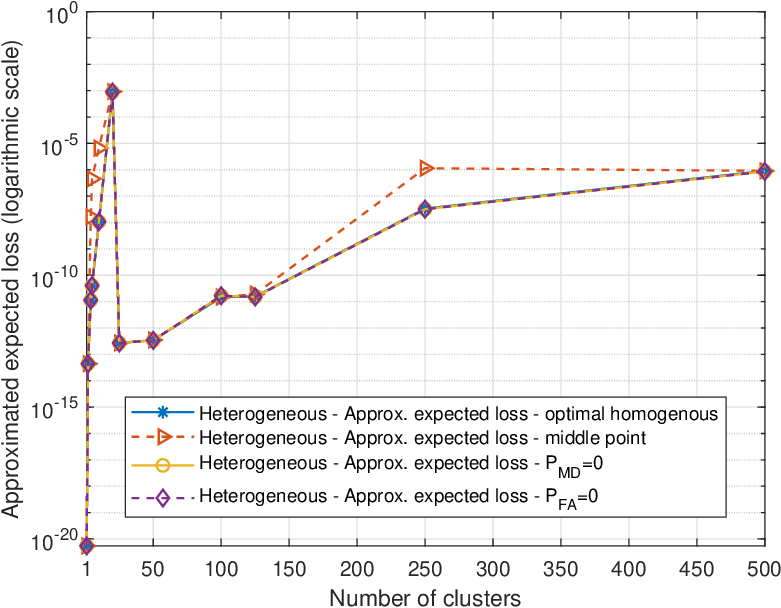}
	\caption{A system with sensor communication probability $p_{\text{com},s_i}=0.4$.}
	\label{Het_compare_init_fig_plot_p_c=0.5}
	 \end{subfigure}
  \vspace{0.25cm}
	 \caption{The expected loss  as a function of the number of the equal sized  clusters, for each of the initial thresholds presented in Section \ref{sec:starting_points_het}. Similarly to Fig.~\ref{fig_Het_compare_init_fig_plot_N_c} the  ``optimal homogeneous" initial threshold which minimizes the expected loss function assuming identical clusters consistently outperforms or is on-par with  the other  candidates.}
\label{fig_Het_compare_init_fig_plot_p_c}
\end{figure}

Figs. \ref{fig_Het_compare_init_fig_plot_N_c}-\ref{fig_Het_compare_init_fig_plot_p_c} evaluate the approximate loss that each of the initial inputs of Algo. \ref{algo:system_heterogeneous} that we present in Section \ref{sec:starting_points_het} yields. Comparing the four initial thresholds for Algo. \ref{algo:system_heterogeneous}, we can see that the first initial threshold that we propose in Section \ref{sec:starting_points_het}, which chooses for each cluster the threshold that minimizes the expected loss function assuming identical clusters, is consistently on-par or  outperforms the other three initial threshold values we propose in Section \ref{sec:starting_points_het}.    

To evaluate the exact performance achieved by  thresholds that are optimized using the approximations that we present in Section \ref{sec:approx_prob_heterogeneous}, we use a homogeneous setup with equal cluster size as a tractable setup for which we can calculate the expected loss exactly. We then compare the exact calculation to its
approximation that is calculated using Eqs. \eqref{eq:upper_cluster_error_FA}-\eqref{eq:upper_FC_error_MD}. 
In the heterogeneous setup we choose the initial threshold $\gamma_j$ for each cluster $\mathcal{C}_j$ using the first initial threshold that we propose in Section\ref{sec:starting_points_het}. In the homogeneous setup we optimize the system by using  Algo. \ref{algo:homonegenous_setup_thoreshold}. Additionally,
in both the heterogeneous setup and the approximate calculation in the homogeneous setup we use the approximate probabilities to approximate  $P_{\text{FA},\mathcal{C}_j}$ and  $P_{\text{MD},\mathcal{C}_j}$ presented in Section \ref{sec:concentration_clusters_FC} if $n_{\mathcal{C}_j}>20$. Additionally,  we use the approximate missed detection and false alarm probabilities to approximate  $P_{\text{FA}}$ and  $P_{\text{MD}}$, i.e., the error probabilities at the FC, presented in Section \ref{sec:concentration_clusters_FC} if $N_c>10$. Otherwise we use exact calculations.

\begin{figure}[t]
 	\centering
	\includegraphics[scale=0.6]{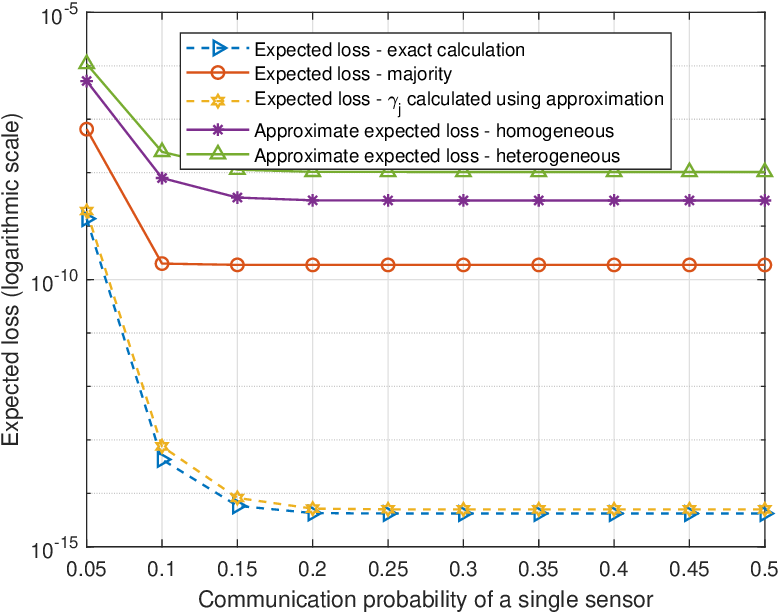}
	\caption{The expected loss function of the communication probability of each sensor for a system with $10$ clusters, each including $50$ sensors. For cloud-cluster architectures we attain a dramatic improvement in performance due to clustering if sensor communication probability to the cloud is at least $0.15$.}
	\label{fig_plot_N_c=10}
\end{figure}

\begin{figure}
 	\centering
	\includegraphics[scale=0.6]{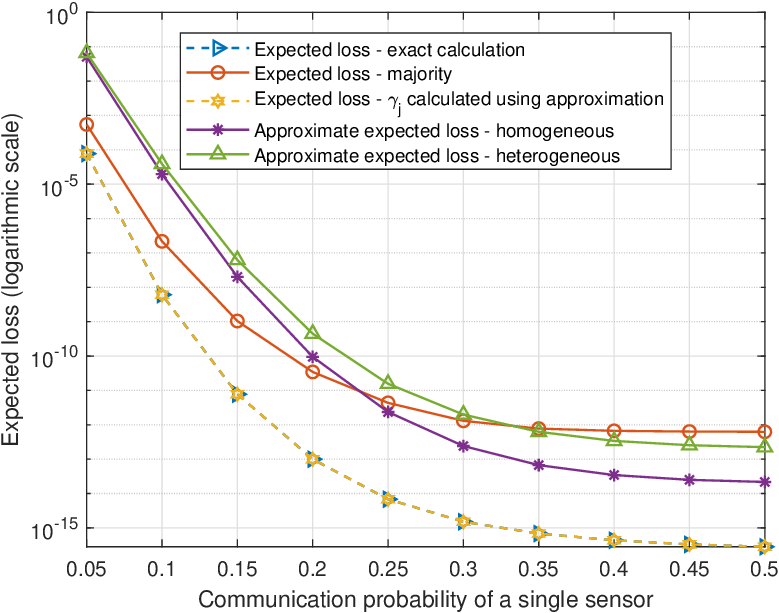}
	\caption{The expected loss function vs. the communication probability of each sensor for a system with $50$ clusters, each including $10$ sensors. For small size clusters, approaching a distributed architecture, higher probability of communication to the cloud is required for better performance.}
	\label{fig_plot_N_c=50}
 \end{figure}

\begin{figure}[t]
 	\centering
	\includegraphics[scale=0.6]{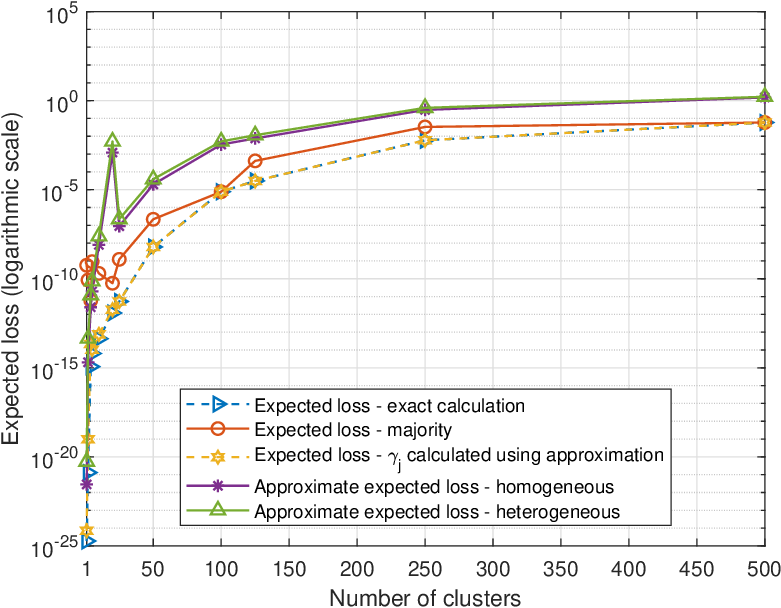}
	\caption{The expected loss function of the number of equal size clusters $N_c$ for $p_{\text{com},s_i}=0.1$. Since connectivity to the FC is low, reducing the number of clusters (more sensors per cluster) increases the chances of communication to the cloud and improves the overall performance.}
	\label{fig_plot_p_c=0.1}
 \end{figure}

\begin{figure}[t]
 	\centering
	\includegraphics[scale=0.6]{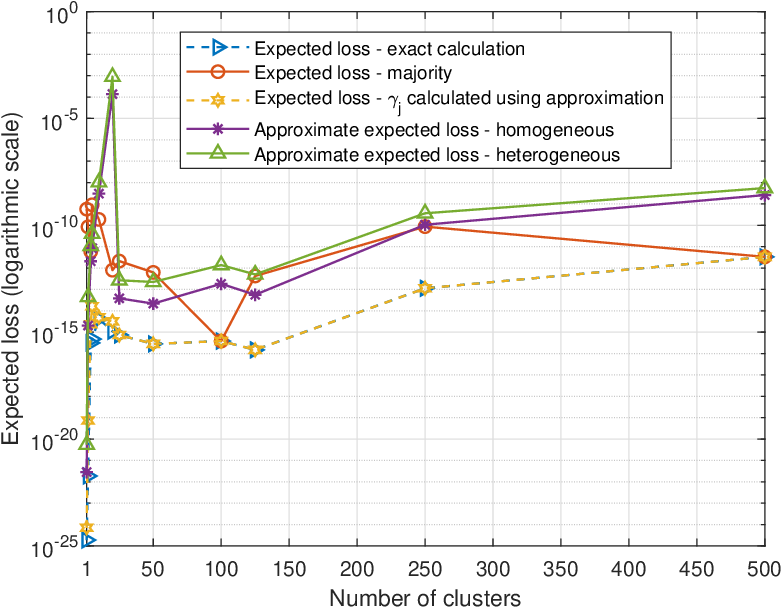}
	\caption{The expected loss function of the number of equal $N_c$ size clusters  for $p_{\text{com},s_i}=0.5$. When connectivity of sensors to the cloud is high, smaller clusters are favored for improving multi-sensor system performance since sensor fusion at the cluster level can be thought of as a form of lossy compression.}
	\label{fig_plot_p_c=0.5}
\end{figure}

Figs. \ref{fig_plot_N_c=10}-\ref{fig_plot_N_c=50} depict the expected loss as a function of the sensor communication probability $p_{\text{com},s}$ for various values of $N_c$ (the number of clusters). 
 Figs. \ref{fig_plot_p_c=0.1}-\ref{fig_plot_p_c=0.5} depict the expected loss as a function  of the number of clusters $N_c$ that comprise the system for various values of sensor communication probabilities $p_{\text{com},s}$. 
Each of the  Figs. \ref{fig_plot_N_c=10}-\ref{fig_plot_p_c=0.5} includes five lines also denoted in the legends. These are defined as: \\
\textbf{Expected loss - exact calculation}: the expected loss of the homogeneous system using exact calculations in Algo. \ref{algo:homonegenous_setup_thoreshold}.\\
\textbf{Expected loss - majority}: the expected loss of the homogeneous system in which each cluster makes a majority rule decision where $\gamma_j = \lfloor n_{\mathcal{C}_j}/2\rfloor+1$. The expected loss is calculated exactly.\\
\textbf{Expected loss - $\gamma_j$ calculated using approximation}: the exact expected loss that the choice $\gamma_j$ yields, where $\gamma_j$ is optimized using the concentration inequalities depicted in Section \ref{sec:concentration_clusters_FC} in Algo. \ref{algo:homonegenous_setup_thoreshold} instead of the exact calculation of the loss function. \\
\textbf{Approximate expected loss - homogeneous}: the approximate expected loss that is calculated using the concentration inequalities depicted in Section \ref{sec:concentration_clusters_FC} in Algo. \ref{algo:homonegenous_setup_thoreshold} instead of the exact calculation of the loss function. \\
\textbf{Approximate expected loss - heterogeneous}: the  approximate expected loss that is calculated using  Algo. \ref{algo:system_heterogeneous} with the first initial threshold that is proposed in Section\ref{sec:starting_points_het}. 

Figs. \ref{fig_plot_N_c=10}-\ref{fig_plot_N_c=50} show that when the number of clusters is large (i.e., each cluster consists of a small number of sensors), the improvement in the performance of a highly connected system compared with that of a sparsely connected system is much more significant than the contrasting scenario of a system with a small number of clusters. Additionally, Figs. \ref{fig_plot_N_c=10}-\ref{fig_plot_N_c=50} confirm that  optimizing the thresholds $\gamma_j$ using concentration inequalities yield an actual expected loss that is on par with that of optimizing $\gamma_j$ using exact calculations. Additionally, Figs. \ref{fig_plot_N_c=10}-\ref{fig_plot_N_c=50} depict the gap between the approximate loss function and the exact one for the homogeneous setup and show that our use of the improved Bennet's inequality results in a good approximation for the expected loss function. Therefore, while the heterogeneous setup is not tractable we can expect that our use of the improved Bennet's inequality results in a good approximation for the expected loss function for the heterogeneous setup as well. Finally, Figs. \ref{fig_plot_N_c=10}-\ref{fig_plot_N_c=50} shows the large gain that optimizing the threshold values provides instead of choosing a majority decision rule.

Figs. \ref{fig_plot_p_c=0.1}-\ref{fig_plot_p_c=0.5} show that when the communication probabilities of sensors to the FC are low,  as in Fig. \ref{fig_plot_p_c=0.1}, there is a monotonic decrease in the  loss function as we decrease the number of clusters in the exact loss function. This is also observed for the approximate loss function with the exception of a small increase when the system is composed of $20$ clusters; the small increase in this case is an artifact resulting from being the first point which approximates \textit{both} the cluster level and the FC error probabilities.
When the communication probabilities of sensors to the FC are higher,  as in Fig. \ref{fig_plot_p_c=0.5}, clustering may actually increase the expected loss. This follows because of the single bit compression that occurs in the clusters' single bit decisions. Note that in this scenario the increase around the point $N_c=20$ is much sharper due to the increase in the exact expected loss and utilizing approximates of both the cluster level and the FC error probabilities.
Fig. \ref{fig_plot_p_c=0.5} exhibits a trade-off between the error probabilities of the decisions in clusters and that of the FC. Increasing the number of clusters reduces the number of measurements that the clusters use to make their decisions, and also reduces the communication probability to the FC since clusters include fewer sensors and thus reduced the chances of seeing an opportunity to access the cloud. However, if the communication probability is high, increasing the number of clusters can result in the FC having more measurements to rely on upon making its final decision.

\section{Conclusion}\label{sec:conclusion}
We consider multi-sensor systems that operate in environments where cloud connectivity is available intermittently. We provide an analytical study of the tradeoffs between different information exchange architectures to support an event detection task. Our results show that if cloud connectivity is reliable, directing sensors to share their sensed values to the cloud for event detection at a centralized fusion center will always perform best. However, in the more likely scenario where cloud connectivity is intermittent, clustering sensors into local neighborhoods where their sensed values are processed and then sent to the cloud during sporadic communication opportunities performs best.  In particular, our results give insight into the optimal cluster sizes needed to achieve minimum detection loss at the cloud even in the face of noisy sensor data and intermittent communication. Future work can use the results presented here to optimize the locations of sensors such that they attain the recommended cluster sizes for best detection performance over the environment.

\appendices
\section{Primer on Concentration Inequalities}\label{append:primer_concentration}

This appendix provides a primer on key concentration inequality results that we will use for the development of our analysis. Since we consider a heterogeneous setup in which the false alarm and missed detection probabilities may vary, we cannot use the concentration inequality \cite{ARRATIA1989} for the binomial distribution. Instead we use an improved Bennett's  inequality which is known to outperform both Bernstein and Hoeffding's inequalities, as well as the Bennet's inequality \cite{doi:10.1080/01621459.1962.10482149}.  
\begin{theorem}[Bennet's inequality \cite{doi:10.1080/01621459.1962.10482149}]
	Let $x_1\ldots,x_n$ be independent random variables and $E(x_i)=0$, $E(x_i^2)=\sigma_i^2$ and $|x_i|<M$ almost surely. Then,  
	\[\Pr\left(\sum_{i=1}^nx_i\geq \alpha\right)\leq \exp\left(-\frac{n\sigma^2}{M^2}h\left(
	\frac{\alpha M}{n\sigma^2}\right)\right),\] 
    for any $0\leq \alpha<nM$,
	where $h(x)=(1+x)\ln(1+x)-x$ and $n\sigma^2=\sum_{i=1}^n\sigma_i^2$.
\end{theorem}

\begin{theorem}[The improved Bennet's inequality \cite{doi:10.1080/03610926.2017.1367818}]\label{theorem:improved_bennett}
	Assume that $x_1\ldots,x_n$ are independent random variables and $E(x_i)=0$, $E(x_i^2)=\sigma_i^2$ and $|x_i|<M$ almost surely.  Additionally, let $\sigma^2=\frac{1}{n}\sum_{i=1}^n\sigma_i^2$ and
	\begin{align}\label{eq:A_B_def}
	A &= \frac{M^2}{\sigma^2}+\frac{nM}{\alpha}-1,\nonumber\\
	B &= \frac{nM}{\alpha} -1,\nonumber\\ 
    \Lambda &=A-W(Be^A),
	\end{align}
	where $W(\cdot)$ is the Lambert $W$ function.
	Denote 
 \begin{flalign}
&U(n,\alpha,M,\sigma^2)\triangleq\nonumber\\
&\hspace{1cm}\exp\left[-\frac{\Lambda \alpha}{M}+n\ln\left(1+\frac{\sigma^2}{M^2}\left(e^{\Lambda}-1-\Lambda\right)\right)\right].
\end{flalign}
	Then, for any $0\leq \alpha<nM$
	\[\Pr\left(\sum_{i=1}^nx_i\geq \alpha\right) \leq U(n,\alpha,M,\sigma^2).
	\]
\end{theorem}

\section{}\label{append:proof_upper_cluster_error_FA_MD}
\begin{IEEEproof}[Proof of Proposition \ref{prop:upper_cluster_error_FA}]
Recall that 
$\tilde{y}_{i} = w_{1,s_{i}}y_{i}-w_{0,s_{i}}(1-y_{i})$.
We can upper bound the false alarm probability \eqref{P_errors_inside} by
\begin{flalign*}
&P_{\text{FA},\mathcal{C}_j} \leq \nonumber\\
&\Pr\left(\sum_{i:s_i\in\mathcal{C}_{j}}\left[\tilde{y}_{i}-E\left(\tilde{y}_i|\mathcal{H}_0\right)\right]\geq \gamma_j-\sum_{i:s_i\in\mathcal{C}_{j}}E\left(\tilde{y}_{i}|\mathcal{H}_0\right)\bigg|\mathcal{H}_0\right).
\end{flalign*}
Furthermore, 
	\begin{flalign} 
	 E\left(\tilde{y}_{i}|\mathcal{H}_0\right)&=P_{\text{FA},s_{i}}w_{1,s_{i}}-(1-P_{\text{FA},s_{i}})w_{0,s_{i}}, \text{ and }\nonumber\\
	E(\tilde{y}_i^2|\mathcal{H}_0) &= P_{\text{FA},s_{i}}w_{1,s_{i}}^2+(1-P_{\text{FA},s_{i}})w_{0,s_{i}}^2.
	\end{flalign}
	It follows that
	\begin{flalign}
	\sigma_{\text{FA},s_{i}}^2&\triangleq\text{var}\left(\tilde{y}_{i}-E\left(\tilde{y}_{i}|\mathcal{H}_0\right)|\mathcal{H}_0\right)\nonumber\\
 &=\text{var}\left(\tilde{y}_{i}|\mathcal{H}_0\right)\nonumber\\
 &=P_{\text{FA},s_{i}}(1-P_{\text{FA},s_{i}})(w_{1,s_{i}}+w_{0,s_{i}})^2.
	\end{flalign}

Now, we can use Theorem \ref{theorem:improved_bennett} to upper bound the false alarm probability of the decision of cluster $j$ by substituting
\begin{flalign*}
x_i&=\tilde{y}_i-E\left(\tilde{y}_{i}|\mathcal{H}_0\right)\nonumber\\
&=\tilde{y}_i-P_{\text{FA},s_{i}}w_{1,s_{i}}+(1-P_{\text{FA},s_{i}})w_{0,s_{i}},\nonumber\\
\alpha_{\text{FA},j}
&=\gamma_j-\sum_{i:s_i\in\mathcal{C}_{j}}E(\tilde{y}_{i}|\mathcal{H}_0)\nonumber\\
&=\gamma_j-\sum_{i:s_i\in\mathcal{C}_{j}}(P_{\text{FA},s_{i}}w_{1,s_{i}}-(1-P_{\text{FA},s_{i}})w_{0,s_{i}}).
\end{flalign*}
Recall that $P_{\text{FA},s_{i}}\in(0,0.5)$. It follows that
$\sigma_{\text{FA},j}^2
=\frac{1}{n_{\mathcal{C}_j}}\sum_{i:s_i\in\mathcal{C}_{j}}P_{\text{FA},s_{i}}(1-P_{\text{FA},s_{i}})(w_{1,s_{i}}+w_{0,s_{i}})^2$,
and $M_{\text{FA},j} = \max_{i:s_i\in\mathcal{C}_{j}}\{m_{\text{FA},i}\}$
where
\begin{flalign*}
m_{\text{FA},i}  &=\max\left\{\left\lvert w_{1,s_{i}} - E\left(\tilde{y}_{i}|\mathcal{H}_0\right)\right\rvert,\left\lvert w_{0,s_{i}} + E\left(\tilde{y}_{i}|\mathcal{H}_0\right)\right\rvert\right\}\nonumber\\
&=(1-P_{\text{FA},s_{i}})(w_{1,s_{i}}+w_{0,s_{i}}).
\end{flalign*}
We denote the resulting constants defined in Theorem \ref{theorem:improved_bennett} by $A_{\text{FA},j}$, $B_{\text{FA},j}$ and $\Lambda_{\text{FA},j}$.  
Thus, by the improved Bennett's inequality, we have that
$
P_{\text{FA},\mathcal{C}_j} \leq U\left(n_{\mathcal{C}_j},\alpha_{\text{FA},j},M_{\text{FA},j},\sigma_{\text{FA},j}^2\right)$,
for every $\gamma_j$ such that 
$0\leq \gamma_j-\sum_{i:s_i\in\mathcal{C}_{j}}(P_{\text{FA},s_{i}}w_{1,s_{i}}-(1-P_{\text{FA},s_{i}})w_{0,s_{i}})<n_{\mathcal{C}_j}\cdot M_{\text{FA},j}$.
\end{IEEEproof}

\begin{IEEEproof}[Proof of Proposition \ref{prop:upper_cluster_error_MD}]
Similarly to the proof of Proposition \ref{prop:upper_cluster_error_FA}, we can use Theorem \ref{theorem:improved_bennett} to upper bound the missed detection probability of cluster $j$. Recall that 
$\tilde{y}_{i} = w_{1,s_{i}}y_{i}-w_{0,s_{i}}(1-y_{i})$.
We upper bound the missed detection probability, $P_{\text{MD},\mathcal{C}_j}$, in \eqref{P_errors_inside} as follows
\begin{flalign*}
&P_{\text{MD},\mathcal{C}_j}\leq\nonumber\\
&\Pr\left(\sum_{i:s_i\in\mathcal{C}_{j}}\left[E\left(\tilde{y}_{i}|\mathcal{H}_1\right)-\tilde{y}_{i}\right]\geq\sum_{i:s_i\in\mathcal{C}_{j}}E\left(\tilde{y}_{i}|\mathcal{H}_1\right)-\gamma_j\bigg|\mathcal{H}_1\right).
\end{flalign*}
Furthermore,
\begin{flalign*}
E\left(\tilde{y}_{i}|\mathcal{H}_1\right)&=(1-P_{\text{MD},s_{i}})w_{1,s_{i}}-P_{\text{MD},s_{i}}w_{0,s_{i}}, \text{ and }\nonumber\\
	E(\tilde{y}_{i}^2|\mathcal{H}_1) &= (1-P_{\text{MD},s_{i}})w_{1,s_{i}}^2+P_{\text{MD},s_{i}}w_{0,s_{i}}^2.
\end{flalign*}
It follows that
\begin{flalign*}
	\sigma_{\text{MD},s_{i}}^2&\triangleq\text{var}\left(E\left(\tilde{y}_{i}|\mathcal{H}_1\right)-\tilde{y}_{i}|\mathcal{H}_1\right)\nonumber\\
 &=\text{var}\left(\tilde{y}_i|\mathcal{H}_1\right)\nonumber\\
 &=P_{\text{MD},s_{i}}(1-P_{\text{MD},s_{i}})(w_{1,s_{i}}+w_{0,s_{i}})^2.
\end{flalign*}
Now, we use Theorem \ref{theorem:improved_bennett} to upper bound the missed detection probability of the decision of cluster $j$ by substituting
\begin{flalign*}
x_i&=E\left(\tilde{y}_{i}|\mathcal{H}_1\right)-\tilde{y}_{i}\nonumber\\
&=(1-P_{\text{MD},s_{i}})w_{1,s_{i}}-P_{\text{MD},s_{i}}w_{0,s_{i}}-\tilde{y}_{i},\\
\alpha_{\text{MD},j}
&=\sum_{i:s_i\in\mathcal{C}_{j}}E(\tilde{y}_{i}|\mathcal{H}_1)-\gamma_j\nonumber\\
&=\sum_{i:s_i\in\mathcal{C}_{j}}((1-P_{\text{MD},s_{i}})w_{1,s_{i}}-P_{\text{MD},s_{i}}w_{0,s_{i}})-\gamma_j.
\end{flalign*}
Recall that $P_{\text{MD},s_{i}}\in(0,0.5)$. It follows that
$\sigma_{\text{MD},j}^2
=\frac{1}{n_{\mathcal{C}_j}}\sum_{i:s_i\in\mathcal{C}_{j}}P_{\text{MD},s_{i}}(1-P_{\text{MD},s_{i}})(w_{1,s_{i}}+w_{0,s_{i}})^2$,
and $M_{\text{MD},j} = \max_{i:s_i\in\mathcal{C}_{j}}\{m_{\text{MD},i}\}$, where
\begin{flalign*}
m_{\text{MD},i} &=\max\left\{\left\lvert w_{1,s_{i}} - E\left(\tilde{y}_{i}|\mathcal{H}_1\right)\right\rvert,\left\lvert w_{0,s_{i}} + E\left(\tilde{y}_{i}|\mathcal{H}_1\right)\right\rvert\right\}\nonumber\\
&=(1-P_{\text{MD},s_{i}})(w_{1,s_{i}}+w_{0,s_{i}}).
\end{flalign*}
We denote the resulting constants defined in Theorem \ref{theorem:improved_bennett} by $A_{\text{MD},j}$, $B_{\text{MD},j}$ and $\Lambda_{\text{MD},j}$. 
By the improved Bennet's inequality we have that
$
P_{\text{MD},j} \leq U\left(n_{\mathcal{C}_j},\alpha_{\text{MD},j},M_{\text{MD},j},\sigma_{\text{MD},j}^2\right)
$, 
for every $\gamma_j$ such that  
$0\leq \sum_{i:s_i\in\mathcal{C}_{j}}((1-P_{\text{MD},s_{i}})w_{1,s_{i}}-P_{\text{MD},s_{i}}w_{0,s_{i}})-\gamma_j<n_{\mathcal{C}_j}M_{\text{MD},j}$.
\end{IEEEproof}

 \section{}\label{append:proof_upper_FC_error_FA_MD}
\begin{IEEEproof}[Proof of Proposition \ref{prop:upper_FC_error_FA}] Denote 
    \[\tilde{z}_{j} = \tau_j\left[w_{1,\mathcal{C}_j}z_j-w_{0,\mathcal{C}_j}(1-z_j)\right].\]
We rewrite the false alarm probability in \eqref{eq:error:prob_all_cluster2} as
	\begin{flalign*}
P_{\text{FA}} &=\Pr\Bigg(\sum_{j=1}^{N_c}\left[\tilde{z}_j-E(\tilde{z}_j|\mathcal{H}_0)\right]\geq\gamma-\sum_{j=1}^{N_c}E(\tilde{z}_j|\mathcal{H}_0)\bigg|\mathcal{H}_0\Bigg). 
\end{flalign*}
By the law of total expectation on $\tau_j$,
	\begin{flalign*} 
	 E\left(\tilde{z}_{j}|\mathcal{H}_0\right)&=p_{\text{com},\mathcal{C}_j} \left[P_{\text{FA},\mathcal{C}_{j}}w_{1,\mathcal{C}_{j}}-(1-P_{\text{FA},\mathcal{C}_{j}})w_{0,\mathcal{C}_{j}}\right] \nonumber\\
  &\triangleq{E_{0,j}}, \nonumber\\
	E(\tilde{z}_j^2|\mathcal{H}_0) &=p_{\text{com},\mathcal{C}_j} \left[ P_{\text{FA},\mathcal{C}_{j}}w_{1,\mathcal{C}_{j}}^2+(1-P_{\text{FA},\mathcal{C}_{j}})w_{0,\mathcal{C}_{j}}^2\right].
	\end{flalign*}
		It follows that
	\begin{flalign*}
	\sigma_{\text{FA},\mathcal{C}_{j}}^2&\triangleq\text{var}\left(\tilde{z}_{j}-E\left(\tilde{z}_{j}|\mathcal{H}_0\right)|\mathcal{H}_0\right)\nonumber\\
 &=\text{var}\left(\tilde{z}_{j}|\mathcal{H}_0\right)\nonumber\\
 &=p_{\text{com},\mathcal{C}_j} \left[ P_{\text{FA},\mathcal{C}_{j}}w_{1,\mathcal{C}_{j}}^2+(1-P_{\text{FA},\mathcal{C}_{j}})w_{0,\mathcal{C}_{j}}^2\right] -E_{0,j}^2.
	\end{flalign*} 
We  use Theorem \ref{theorem:improved_bennett} to upper bound the false alarm probability of the final decision of the FC by substituting $j$ with $i$ in Theorem \ref{theorem:improved_bennett} and 
\begin{flalign*}
x_j&=\tilde{z}_j-E\left(\tilde{z}_j|\mathcal{H}_0\right)=\tilde{z}_j-E_{0,j},\nonumber\\
\alpha_{\text{FA}}
&=\gamma-\sum_{j=1}^{N_c}E(\tilde{z}_j|\mathcal{H}_0)=\gamma-\sum_{j=1}^{N_c}E_{0,j}.
\end{flalign*}
In this case, 
\begin{flalign*}
&\sigma_{\text{FA}}^2=\nonumber\\
&\frac{1}{N_c}\sum_{j=1}^{N_c}\left[p_{\text{com},\mathcal{C}_j} \left( P_{\text{FA},\mathcal{C}_{j}}w_{1,\mathcal{C}_{j}}^2+(1-P_{\text{FA},\mathcal{C}_{j}})w_{0,\mathcal{C}_{j}}^2\right) -E_{0,j}^2\right].
\end{flalign*}
Additionally, $M_{\text{FA}} = \max_{j\in[1:N_c]}\{m_{\text{FA},j}\}$, where
\begin{flalign*}
m_{\text{FA},j} 
&=\max\left\{\left\lvert w_{1,\mathcal{C}_j} - E\left(\tilde{z}_j|\mathcal{H}_0\right)\right\rvert,\left\lvert w_{0,\mathcal{C}_j} + E\left(\tilde{z}_j|\mathcal{H}_0\right)\right\rvert\right\}\nonumber\\
&=\max\{|w_{1,\mathcal{C}_j}-E_{0,j}|,|w_{0,\mathcal{C}_j}+E_{0,j}|\}.
\end{flalign*}
We denote the resulting constants defined in Theorem \ref{theorem:improved_bennett} by $A_{\text{FA}}$, $B_{\text{FA}}$ and $\Lambda_{\text{FA}}$.  
It follows from the improved Bennett's inequality  that $P_{\text{FA}} \leq U\left(N_c,\alpha_{\text{FA}},M_{\text{FA}},\sigma_{\text{FA}}^2\right)$, for every $\gamma$ such that  
$0\leq \gamma-\sum_{j=1}^{N_c}E_{0,j}<N_c\cdot M_{\text{FA}}$.
\end{IEEEproof}

\begin{IEEEproof}[Proof of Proposition \ref{prop:upper_FC_error_MD}]
Similarly to the proof of Proposition \ref{prop:upper_FC_error_FA}, we can use Theorem \ref{theorem:improved_bennett} to upper bound the missed detection probability of the final decision of the FC. 
Recall that $
    \tilde{z}_{j} = \tau_j\left[w_{1,\mathcal{C}_j}z_j-w_{0,\mathcal{C}_j}(1-z_j)\right]$.	
    We can rewrite the missed detection probability in \eqref{eq:error:prob_all_cluster2} as
	\begin{flalign*}
P_{\text{MD}}&=\Pr\left(\sum_{j=1}^{N_c}\left[E\left(\tilde{z}_{j}|\mathcal{H}_1\right)-\tilde{z}_{j}\right]>\sum_{j=1}^{N_c}E\left(\tilde{z}_{j}|\mathcal{H}_1\right)-\gamma_j\bigg|\mathcal{H}_1\right). 
\end{flalign*}
By the law of total expectation on $\tau_j$,
	\begin{flalign*} 
	 E\left(\tilde{z}_{j}|\mathcal{H}_1\right)&=p_{\text{com},\mathcal{C}_j}\left[ (1-P_{\text{MD},\mathcal{C}_{j}})w_{1,\mathcal{C}_{j}}-P_{\text{MD},\mathcal{C}_{j}}w_{0,\mathcal{C}_{j}}\right]\nonumber\\
  &\triangleq E_{1,j},\nonumber\\
	E(\tilde{z}_{j}^2|\mathcal{H}_1) &=p_{\text{com},\mathcal{C}_j} \left[ (1-P_{\text{MD},\mathcal{C}_{j}})w_{1,\mathcal{C}_{j}}^2+P_{\text{MD},\mathcal{C}_{j}}w_{0,\mathcal{C}_{j}}^2\right].
	\end{flalign*}
		It follows that,
	\begin{flalign*}
	\sigma_{\text{MD},\mathcal{C}_{j}}^2&\triangleq\text{var}\left(E\left(\tilde{z}_{j}|\mathcal{H}_1\right)-\tilde{z}_{j}|\mathcal{H}_1\right)\nonumber\\
 &=\text{var}\left(\tilde{z}_j|\mathcal{H}_1\right)\nonumber\\
 &=p_{\text{com},\mathcal{C}_j} \left[ (1-P_{\text{MD},\mathcal{C}_{j}})w_{1,\mathcal{C}_{j}}^2+P_{\text{MD},\mathcal{C}_{j}}w_{0,\mathcal{C}_{j}}^2\right] -E_{1,j}^2.
	\end{flalign*} 
We use Theorem \ref{theorem:improved_bennett} we upper bound the missed detection probability of the final decision of the FC by substituting $j$ with $i$ in Theorem \ref{theorem:improved_bennett} and  
\begin{flalign*}
x_j&=E\left(\tilde{z}_j|\mathcal{H}_1\right)-\tilde{z}_j=E_{1,j}-\tilde{z}_j,\nonumber\\ 
\alpha_{\text{MD}}&=\sum_{j=1}^{N_c}E(\tilde{z}_j|\mathcal{H}_1)-\gamma=\sum_{j=1}^{N_c}E_{1,j}-\gamma.
\end{flalign*}
In this case, 
\begin{flalign}
&\sigma_{\text{MD}}^2=\nonumber\\
&\frac{1}{N_c}\sum_{j=1}^{N_c}\Big[p_{\text{com},\mathcal{C}_j} \left( (1-P_{\text{MD},\mathcal{C}_{j}})w_{1,\mathcal{C}_{j}}^2+P_{\text{MD},\mathcal{C}_{j}}w_{0,\mathcal{C}_{j}}^2\right)-E_{1,j}^2\Big],
\end{flalign}
and $M_{\text{MD}} = \max_{j\in[1:N_c]}\{m_{\text{MD},j}\}$, where
\begin{flalign*}
m_{\text{MD},j} &= 
\max\left\{\left\lvert w_{1,\mathcal{C}_j} - E\left(\tilde{z}_j|\mathcal{H}_1\right)\right\rvert,\left\lvert w_{0,\mathcal{C}_j} + E\left(\tilde{z}_j|\mathcal{H}_1\right)\right\rvert\right\}\nonumber\\
&=\max\big\{|w_{1,\mathcal{C}_j}-E_{1,j}|,|w_{0,\mathcal{C}_j}+E_{1,j}|\big\}.
\end{flalign*}
We denote the resulting constants defined in Theorem \ref{theorem:improved_bennett} by $A_{\text{MD}}$, $B_{\text{MD}}$ and $\Lambda_{\text{MD}}$. 
By the improved Bennet's inequality we have that $P_{\text{MD}} \leq U\left(N_c,\alpha_{\text{MD}},M_{\text{MD}},\sigma_{\text{MD}}^2\right)$, for every $\gamma$ such that  
$0\leq \sum_{j=1}^{N_c}E_{1,j}-\gamma<N_c\cdot M_{\text{MD}}$.
\end{IEEEproof}

\bibliographystyle{IEEEtran}

\end{document}